# Near-real-time data captured record decline in global CO2 emissions due to COVID-19

Zhu Liu[1*†], Philippe Ciais[2†], Zhu Deng[1†], Ruixue Lei[3], Steven J. Davis[4], Sha Feng[4], Bo Zheng[2], Duo Cui[1], Xinyu Dou[1], Pan He[1], Biqing Zhu[1], Chenxi Lu[1], Piyu Ke[1], Taochun Sun[1], Yuan Wang[5,6], Xu Yue[7], Yilong Wang[8], Yadong Lei[8], Hao Zhou[9], Zhaonan Cai[10], Yuhui Wu[11], Runtao Guo[12], Tingxuan Han[13], Jinjun Xue[14, 15, 16], Olivier Boucher[17], Eulalie Boucher[2], Frederic Chevallier[2], Yimin Wei[18], Haiwang Zhong[19], Chongqing Kang[19], Ning Zhang[20], Bin Chen[21], Fengming Xi[22], François Marie[2], Qiang Zhang[1], Dabo Guan[1], Peng Gong[1], Daniel M. Kammen[23], Kebin He[11], Hans Joachim Schellnhuber[24]

[1] Department of Earth System Science, Tsinghua University, Beijing 100084, China.

[2] Laboratoire des Sciences du Climate et de l'Environnement LSCE, Orme de Merisiers 91191 Gif-sur-Yvette, France

[3] Department of Meteorology and Atmospheric Science, The Pennsylvania State University, University Park, PA 16802

[4] Department of Earth System Science, University of California, Irvine, 3232 Croul Hall, Irvine, CA 92697-3100, USA

[5] Division of Geological and Planetary Sciences, California Institute of Technology, Pasadena, CA, USA

[6] Jet Propulsion Laboratory, California Institute of Technology, Pasadena, CA, USA

[7] Jiangsu Key Laboratory of Atmospheric Environment Monitoring and Pollution Control, Collaborative Innovation Center of Atmospheric Environment and Equipment Technology, School of Environmental Science and Engineering, Nanjing University of Information Science & Technology (NUIST), Nanjing 210044, China

[8] Key Laboratory of Land Surface Pattern and Simulation, Institute of Geographical Sciences and Natural Resources Research, Chinese Academy of Sciences, Beijing, China

[9] Climate Change Research Center, Institute of Atmospheric Physics, Chinese Academy of Sciences, Beijing 100029, China

[10] Key Laboratory of Middle Atmosphere and Global Environment Observation, Institute of Atmospheric Physics, Chinese Academy of Sciences, Beijing 100029, China.

[11] School of Environment, Tsinghua University, Beijing 100084, China

[12] School of Mathematical School, Tsinghua University, Beijing 100084, China

[13] Department of Mathematical Sciences, Tsinghua University, Beijing 100084, China

[14] Center of Hubei Cooperative Innovation for Emissions Trading System, Wuhan, China

[15] Faculty of Management and Economics, Kunming University of Science and Technology, 13 Kunming, China

[16] Economic Research Centre of Nagoya University, Furo-cho, Chikusa-ku, Nagoya, Japan

[17] Institute Pierre-Simon Laplace, Sorbonne Université / CNRS, Paris

[18] Center for Energy and Environmental Policy Research, Beijing Institute of Technology, Beijing, China

[19] Department of Electrical Engineering, Tsinghua University, Beijing 100084, China

20 Jinan University, Guangzhou, China

21 Beijing Normal University, Beijing, China

[22] Institute of Applied Ecology, Chinese Academy of Sciences, Shenyang, China

[23] Energy and Resources Group and Goldman School of Public Policy, University of California, Berkeley, CA, USA

[24] Potsdam Institute for Climate Impact Research, 14412 Potsdam, Germany

* Corresponding authors: zhuliu@tsinghua.edu.cn,

† Authors contribute equally

**The considerable cessation of human activities during the COVID-19 pandemic has affected global energy use and $CO_2$ emissions. Here we show the unprecedented decrease in global fossil $CO_2$ emissions from January to April 2020 was of 7.8% (938 Mt $CO_2$ with a ±6.8% of 2-σ uncertainty) when compared with the period last year. In addition other emerging estimates of COVID impacts based on monthly energy supply[1] or estimated parameters[2], this study contributes to another step that constructed the near-real-time daily CO2 emission inventories based on activity from power generation (for 29 countries), industry (for 73 countries), road transportation (for 406 cities), aviation and maritime transportation and commercial and residential sectors emissions (for 206 countries). The estimates distinguished the decline of CO2 due to COVID-19 from the daily, weekly and seasonal variations as well as the holiday events. The COVID-related decreases in CO2 emissions in road transportation (340.4 Mt $CO_2$, -15.5%), power (292.5 Mt $CO_2$, -6.4% compared to 2019), industry (136.2 Mt $CO_2$, -4.4%), aviation (92.8 Mt $CO_2$, -28.9%), residential (43.4 Mt $CO_2$, -2.7%), and international shipping (35.9Mt $CO_2$, -15%). Regionally, decreases in China were the largest and earliest (234.5 Mt $CO_2$, -6.9%), followed by Europe (EU-27 & UK) (138.3 Mt $CO_2$, -12.0%) and the U.S. (162.4 Mt $CO_2$, -9.5%). The declines of CO2 are consistent with regional nitrogen oxides concentrations observed by satellites and ground-based networks, but the calculated signal of emissions decreases (about 1Gt CO2) will have little impacts (less than 0.13ppm by April 30, 2020) on the overserved global CO2 concertation. However, with observed fast CO2 recovery in China and partial re-opening globally, our findings suggest the longer-term effects on CO2 emissions are unknown and should be carefully monitored using multiple measures.**

*[288 words]*

**Introduction**

COVID-19 has caused hundreds of thousands of deaths worldwide since December of 2019, together with large-scale ongoing reductions in human activities and profound effects on different national economies. Industrial production and energy consumption in some countries were reported to decline by up to 30% in just a few weeks[3] as lockdowns were imposed to protect public health. Fossil fuel and cement $CO_2$ emissions are directly linked to human activities. Initial estimates of emissions changes based on a limited sample of power plants and indirect satellite observations of atmospheric pollutants[4,5] have suggested that we may be witnessing the largest drop of emissions since the end of the Second World War. However the detailed inventories of energy and fuel use that have historically been used to assess $CO_2$ emissions are only available with a lag of one or two years[6-12] that result in challenges to assessing the $CO_2$ dynamics due to COVID-19. More recently there are emerging estimates[1,2] that provides pioneering assessment on the impacts, for example, International Energy Agency uses monthly fossil fuel energy demand to conclude a -5% decline of $CO_2$ in Jan-April 2020[1], and another big step by Le Quéré et al.[2] using parameter measures that compile government policies and activity data to estimate the decrease in $CO_2$ emissions during forced confinements, suggesting daily global CO2 emissions decreased by –17% by early April 2020 compared with the mean 2019 levels. Those studies provide initial steps to estimate the extent of COVID impacts on $CO_2$ emission (-4% to -8%) in 2020 and reached profound impacts and attentions from public and academia.

However, there still lacking for quantitative assessment to distinguish the decline of $CO_2$ emissions due to COVID19 from the daily, weekly and seasonal $CO_2$ emission variations as well as the holiday events, and capture the real time dynamics of $CO_2$ emissions to precisely reflect the extent of the $CO_2$ decline due to COVID-19. To do so requires detailed CO2 emission inventories for near real time for year 2020 to date as well as for previous years as the baseline. More importantly, such inventory is critical to monitor the future $CO_2$ dynamics after the forced confinement released gradually since March 2020, in which the emission trends can be various.

Here, we conducted a new step that constructed the near-real-time CO2 emission inventories with temporal resolution to daily in period Jan 2019 to May 2020. Such inventories captured the daily, weekly and seasonal variations within (in 2020) and without COVID-19 (in 2019), as well as impacts by holidays in major countries such as Spring Festival in China, thus the impacts of COVID-19 can be distinguished.

Details of our data sources and analytic methods are provided in the *Methods* section. In summary, we calculated regional $CO_2$ emissions between January and May 2020 and compare them to the calculated regional $CO_2$ emissions in same period in 2019, drawing on hourly datasets of electricity power production and $CO_2$ emissions in 29 countries (including the substantial variations in carbon intensity associated with electricity production), three different indexes of daily vehicle traffic / mobility in 416 cities worldwide, monthly production data for cement, steel and other energy intensive industrial products in 73 countries, daily maritime and aircraft transportation activity data, as well as proxies for the residential and the commercial building emissions (see *Methods* for data sources). Together, these data cover almost all fossil and industry sources of global $CO_2$ emissions, including CO2 emission from cement production that not been considered by IEA assessment[1].

**Results**

The near real time emissions in Jan 1st 2019 to April 30 2020

The near-real-time daily emission inventories show significant decrease of CO2 emission (-7.8%, 937Mt CO2) in Jan 1st from April 30th in 2020 when compared with same period in 2019 (Fig 1). There are also remarkable seasonal (-6%, +8%), weekly(-12%, +12%) and daily (-12%, +14%) min-max variations of CO2 emissions showing in whole year 2019 (Fig 1a) and in January to May 2020 (Fig 1b), mainly due to the heating and cooling demands and difference of activities in different seasons (i.e. heating degree days-HDD[7] and cooling degree days-CDD), weekday and weekend, as well as holiday events. To distinguish the COVID-19 impacts from seasonal variations, we calculated the difference of emission in 2020 from same period in 2019 by comparing the daily emissions in same period (Fig1b), the total difference of 938 Mt $CO_2$ is the largest decline with its decrease rate ever seen on record (Fig 1c), with its half year decrease larger than the total annual decrease (790 Mt CO2) during World War II (Fig 1d). The daily mean emission in Jan-April 2020 (92.0 Mt CO2 per day) is 6% less than daily mean CO2 in same period 2019 (98.0 Mt CO2 per day), the decline is more significant in April (-14.4% in April 2020 compared with 2019), but showing the narrow of decline in May. The cumulated emission in Jan-April 2020 (11127 Mt CO2 for first 121 days in 2020) is 7.8% less than that (12065 Mt $CO_2$ for first 120 days in 2019).

The results showing that the decline slows in late April globally, mainly contributed by the recovery of emission in China and the Europe, along with the release of pandemic in these regions (Fig 1e). The dataset also captured the impact of daily emission decline due to holidays, for example, spring festival in China result in the decline of daily $CO_2$ in China's power sector in 2019 similar to the level during COVID19 pandemic in 2020, and the data also reflect the impact on emission decline from Qingming Holiday(April 5th) and Labor Day Holiday (May 1st) (Fig 1f).

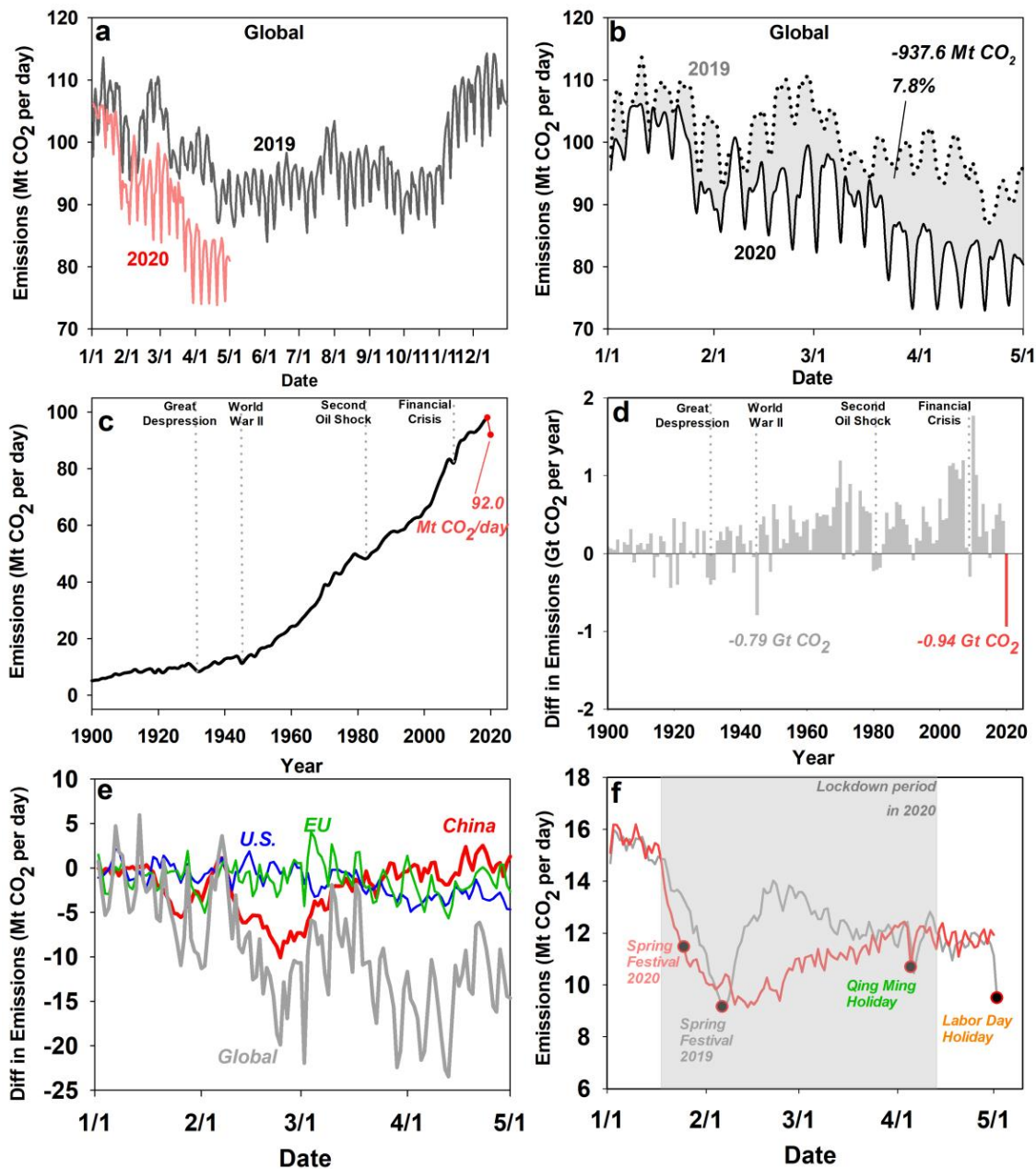

**Figure 1. Daily CO2 emissions from 2019 to April 30th 2020 (a). The difference in 2020 compared with 2019 shows the COVID19 impacts (b), which the decline of 937Mt CO2 is the highest ever in 4 months. The difference of emission in 2020 compared with 2019 is contributed by regional dynamics of CO2 (d). The data also showing decline of emission in Holidays for China's power sector in addition to the impact from COVID-19.**

It is important to note that first months of 2020 were exceptionally warm across much of the northern hemisphere, meaning that 2020 $CO_2$ emissions would have been somewhat lower than the same period in 2019 even without the disruption in economic activities and energy production caused by COVID-19 and related lockdowns. Thus, we considered the impacts on $CO_2$ from COVID-19 by removing the difference of daily $CO_2$ due to temperature variation between 2019 and 2020 in the winter months (January-March), and comparing the $CO_2$ emission in 2020 with the same period in 2019.

The daily CO2 in 2020 show significant decline when compared with 2019, despite the seasonal variations. Figure 2 shows estimated trends in total $CO_2$ emissions globally and in several major regions. Globally, we find a global 7.8% decrease of $CO_2$ emissions during the first four months of 2020 (solid black curve) compared with the same period in 2019 (dashed black curve). In the first quarter of 2020, the most pronounced decline occurred in China, where emissions fell by -9.3%, with substantial but progressively smaller decreases in Europe (EU27 & UK) (-8.4%), the U.S. (-4.7%), Japan (-3.6%), India (-2.5%) and Russia (-2.1%). The large and early drop in Chinese emissions correspond to an early outbreak of COVID-19 and strict lockdown measures, which were relaxed throughout March. However, due to the rapid control of COVID-19 pandemic, the $CO_2$ emission recovered quickly after March, the 2020-2019 difference in China's emissions were much less in March (-8.1%) than in February (-14.6%), and China's CO2 in April 2020 is about 0.8% higher than the $CO_2$ in April 2019. The data also show that emission declines due to China's spring festival reached similar level of the daily emission decline (-10% of daily CO2 decline within Spring festival when compared with the average daily CO2 in February 2019). In other countries, decreases in emissions due to COVID-19 weren't apparent until late February or March, coincident with the onset of lockdowns in different countries, with greater decreases generally observed in March (U.S.:-13.8%, EU27 & UK: -8.1%, India: -16.4%, Brazil:-11.0%, Japan -4.1%) than in February (U.S.: 1.9%, EU27 & UK: -8.4%, India: 6.2%, Brazil: -1.6%, Japan -1.1%), and larger decreases showing in April (U.S.:-25.6%, EU27 & UK: -25.0%, India: -27.9%, Brazil:-26.6 %, Japan -6.7%). So when it comes to the emissions in the first four months of 2020, the largest decrease by -12.0% occurred in Europe (EU27 & UK), with smaller decrease in the U.S. (-9.5%), India (-8.5%), Brazil (-7.0%), China (-6.9%), Japan (-4.3%) and Russia (-3.4%).

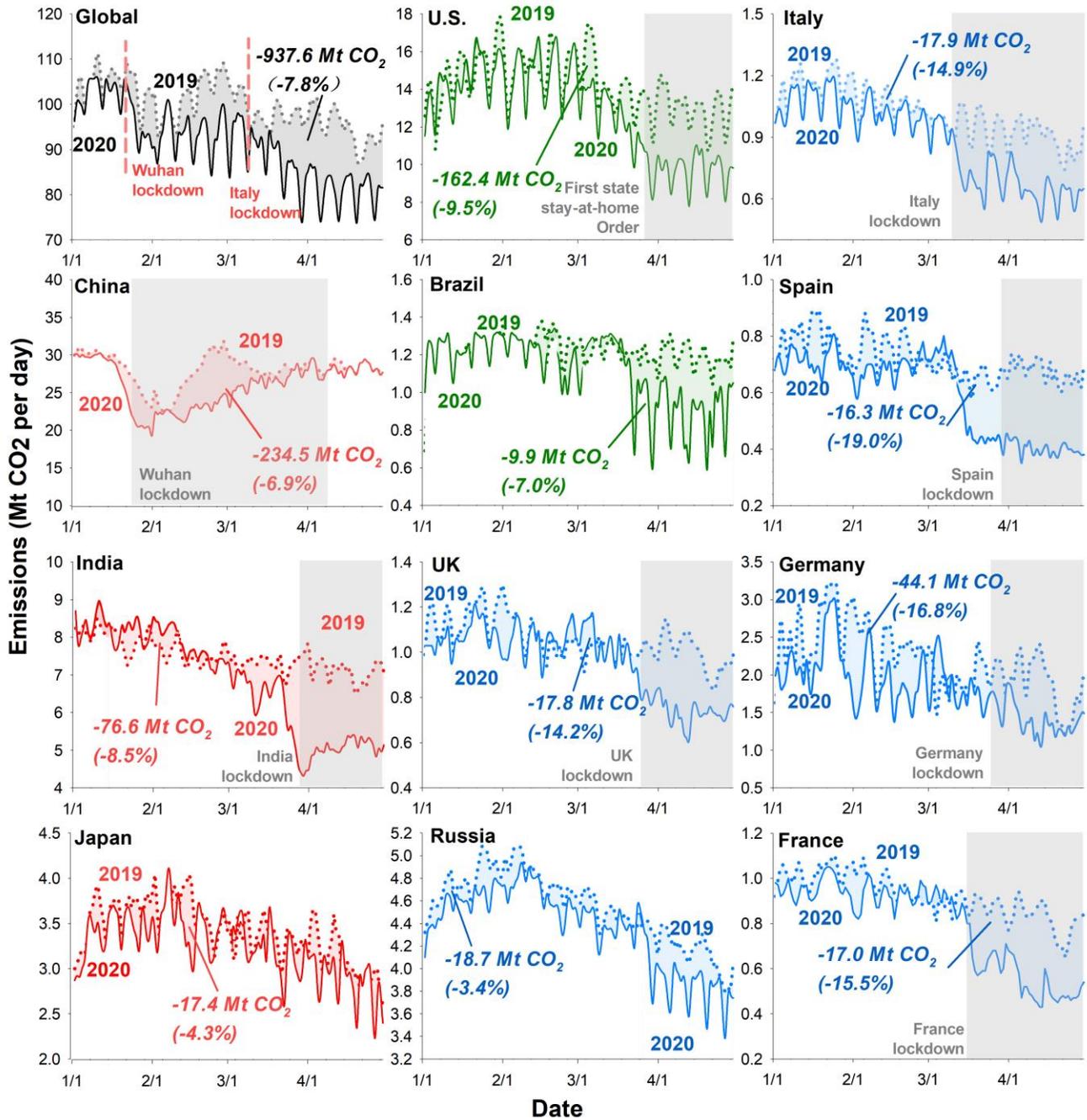

**Figure 2. Daily CO2 emissions in the first quarter of 2019 (dotted line) and 2020 (Solid Line) for the world, U.S., Italy, China, Brazil, Spain, India, UK, Germany, Japan, Russia and France. Different color for countries represents different continents.**

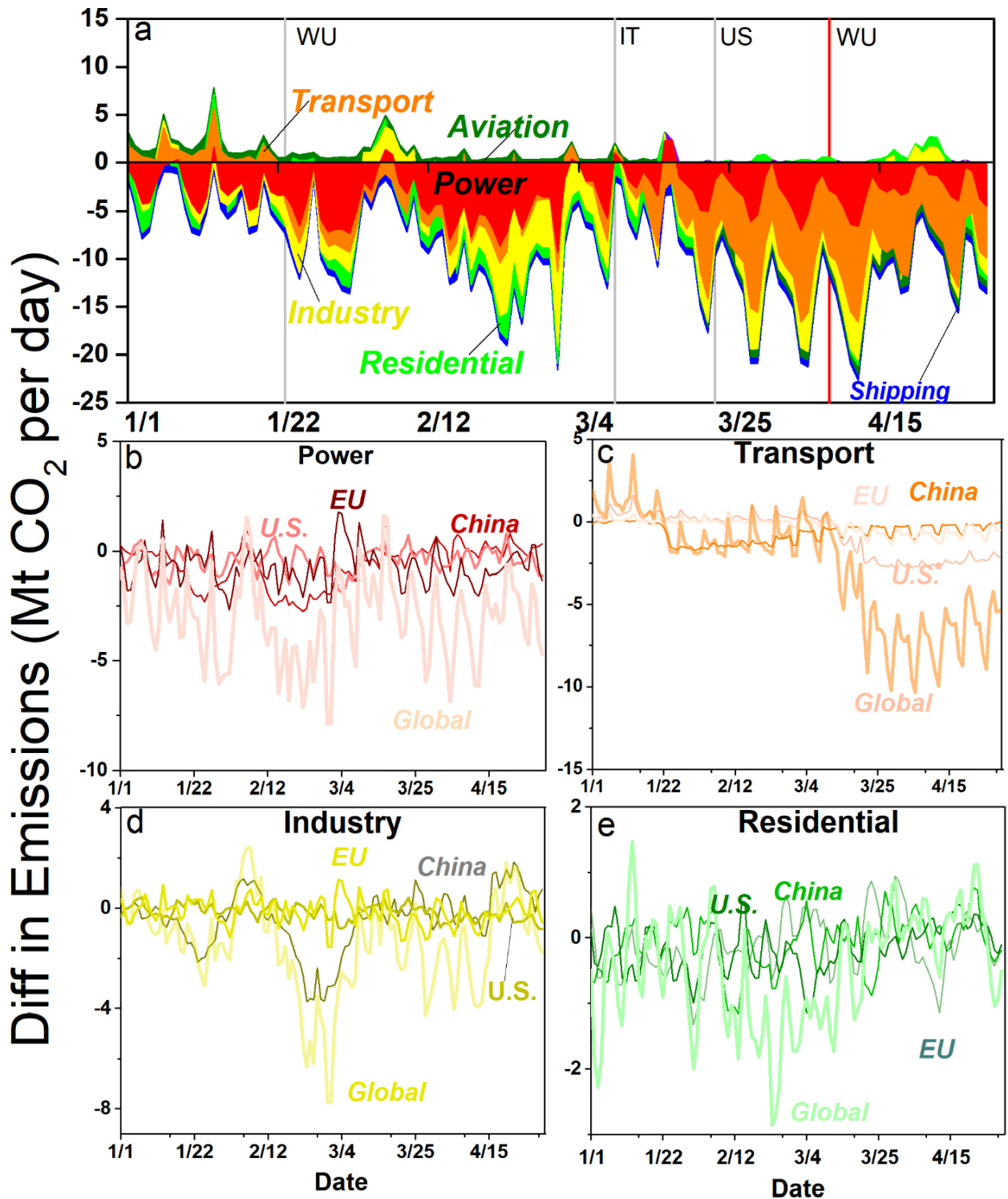

**Figure 3 | (a)** Global fossil fuel and cement CO2 emissions (Seven-days running mean) difference between 2020 and 2019 for different sectors. **(b-e)** Emissions differences for different sectors for different regions. The green lines for the transport sector are for ground

**(mainly road) transport emissions and the light blue line is for aviation emissions changes for the entire globe.**

Figure 3 shows the breakdown of emissions decreases by sector. The largest contributions to the global decrease in emissions come from ground transport (340.0 Mt $CO_2$, 36% of the total; orange in Fig. 3a) and power (-292.5 Mt $CO_2$, 31% of the total first four months decrease; red in Fig. 3a) with decreases from industry sector just slightly less (-136.2 Mt $CO_2$, 15% of the total; yellow in Fig. 3a), and relatively small decreases in residential emissions (-43.4 Mt $CO_2$, 5% of the total; green in Fig. 3a). The rest of the reduction comes from aviation and ships emissions.

### Power Generation

Our estimates of power sector emissions rely on near real time hourly or daily electricity data. Thus, we are able to resolve the effects of weather-driven variations of renewable electricity supply (See *Methods* for temperature adjustment) as well as the increases in natural gas relative to coal for power generation in 2020 in the U.S. that has been caused by very low oil prices. In the winter months (January-March) of 2020, the temperature variation compared to 2019 leads to -0.8% reduction in the power emissions. Figure 1b shows that in the first four months of 2020, global $CO_2$ emissions from the power sector declined by -6.4% (-292.5 Mt $CO_2$), with a small decline in China (-6.0%, -91.1 Mt $CO_2$) and somewhat larger decreases in the U.S. (-7.7%, -43.8 Mt $CO_2$), India (-9.2%, 39.7 Mt $CO_2$) and the EU-27 & UK (-22.5%, -82.0 Mt $CO_2$). Some of the drop in China's power sector emissions are due to prevailing warmer winter temperatures, and the near-zero differences in late January and early February between 2020 and 2019 are because this was when the country's spring festival occurred in 2019 (Fig 1e). The decline of power generation in Ching Ming Festival (April 5th) and the Labor Day holiday (May 1st-5th) are also reflected by the data. Meanwhile, power emissions in Russia and Japan were almost stable in the first four months of 2020 (-2.4%, -7.0 Mt $CO_2$; -2.9%, -5.6 Mt $CO_2$) (See SI Table S2).

### Industry and cement production emissions

Industry emissions from steel, chemicals and other manufactured products from fossil fuel combustion and the cement production process represent on average 29% of the global $CO_2$ emissions, with a much larger share of national emissions in developing countries (39% in China and 33% in India)[13]. We collected data separately for steel, chemicals (based on 8 chemical products) and 26 other industrial products, as well as for cement production in China, for a better attribution of industrial $CO_2$ emissions changes (See SI Table S6). Only emissions from direct fuel consumption and chemical process emissions by the industry sector were considered, electricity related emissions for the industry being already counted in the power generation sector. In the first four months of 2020, global industry emissions fell by -4.4% in most countries, including China (-3.5%, -43.8 Mt $CO_2$), U.S. (-6.4%, -17.1 Mt $CO_2$), EU27 & UK (-7.3%, -15.1 Mt $CO_2$) and India (-7.9%, -22.1 Mt $CO_2$). In China, emissions from steel production (41.6% of national industrial emissions from fuel combustion) remained essentially the same with slight increases in January and February by 1.5% and 5.0%, respectively, and only decreased in March by 1.7% (the production data suggests no difference in emission in April). This rather surprising result demonstrates the core status of steel industry in China's industrial structure. Overall, despite the COVID-19 pandemics, emissions from the steel industry thus resulted 1.4% higher in the first four months of 2020. For the cement industry (22.2% of China's industrial emissions from fuel combustion), estimates based on the official reports from National Bureau of Statistics[14] show a considerable decline of -14.4% in the first four months in 2020, namely -29.5% in January and February combined, -18.3% in March but 3.8% increase in April. The latter recovery substantiated a concurrent recovery in various industrial activities. Emissions from the production of chemicals in China also decreased by 1.5%, while emissions from other industries had fallen by 4.8% in the same time period. In addition, cement production usually entails the decomposition of calcium carbonate in making cement clinker, which produces the so-called process emission. The absolute amount of $CO_2$ abated from such process emission was estimated to be 37.3 Mt in China with the same relative reduction as to the fuel combustion in producing cements.

**Ground transportation emissions**

Emissions from ground transportation (SI Figure S4) were calculated based on TomTom congestion level with daily transportation activity data for 416 global cities in more than 50 countries. Ground transportation (See Methods for data and calculation process) contributes 18% of the world $CO_2$ emissions and decreased dramatically by -15.5% in the first four months (-340 Mt $CO_2$) thus contributing 36.3% of the decline of the emissions from all sectors (Figure 3). In China, as cities started locking down in the last week of January, the average emissions from transport during that month decreased by -18.5%. In February, ground transport emissions dropped abruptly by -53.4%, compared to the same month in 2019. However, the reduction has been shrinking in March (-25.9%) and April (-16.3%) respectively. The emissions from ground transport have dramatically decreased in other countries just after lockdown measures since March. Ground transport emissions in U.S. and India dropped by -22.8% and -25.9% in March since the lockdown measures, and continued to decrease by -50.0% and -65.6% in April respectively. Emissions in European countries dropped by -16.6% and -32.1% in March and April, with France, Spain and Italy showing the largest reductions of -15.8%, -14.2% and 13.9% respectively in the first four month in 2020 (SI Figure S4).

**Aviation and ships emissions**

Emissions from global aviation decreased by -28.9% during the first four months (-92.8 Mt $CO_2$), among which those from international aviation decreased by -63.6 Mt $CO_2$ (SI Figure S5). Domestic aircraft emissions are included in our global estimates, but only international aviation emissions attributed to different countries (See Methods for data and calculation process). The total number of flights and global aviation emissions shows two consecutive decreases, one by the end of January in Asia and another since the middle of March in the rest of the world. Emissions declined sharply after mid-March, coincident with travel bans and lock-down measures. $CO_2$ emissions from international shipping (see Methods for data and calculation process) decreased by -15.0% (-32.6 Mt $CO_2$) in the first four months.

**Commercial and Residential buildings**

We used atmospheric ERA5 temperature data[15] to interpret the consumption variability and accounted for the weekly cycle and the occurrence of holidays that differ among countries and from year to year. The emissions from fuel use (oil and gas) in commercial and residential buildings was estimated using population-weighted heating degree days by the ERA515 reanalysis of 2 meters air temperature for 206 countries. We found that the global heating demand in the first four months declined by -2.7% compared to 2019, owing to the abnormal warm northern-hemisphere winter[16], resulting in the decreased of emissions. The results indicate that the residential consumption is mostly driven by the temperature, and that there was no significant change during the lockdown period (i.e. no significant change with regards to the year-to-year variability when the contribution of temperature variability was removed).

To verify such estimation, we analyzed the natural gas consumption for commercial and residential buildings for 6 countries (France, Italy, Great Britain, Belgium, Netherlands and Spain) of Western Europe for which daily data was available from national operators. The data were converted to CO2 emissions using emission factors that account for gas quality. For these 6 countries, these data converted to CO2 emissions were used after rescaling to annual national residential emissions from EDGAR for the year 2019 to account for additional use of liquid and solid fuels used in the residential sector, which is counted in EDGAR but not covered by our natural gas consumption data. Nevertheless, for the analyzed countries, the industry data show a decrease of the demand by $\approx$20% for a few weeks after the start of the lockdown. Note that the industry consumption is a small fraction of that of the residential sector, so that the lockdown has had an overall small impact on the gas use for these countries. These results support the assumption of the method used for all other countries that the gas consumption was little affected by the lockdown and that the gas consumption decrease or increase during the period was mostly the result of variations of temperatures.

**Observation and verification from air quality data**

Our estimates of decreases in fossil and industry $CO_2$ emissions (See Methods and SI Table S9) are consistent with observed changes in nitrogen dioxide ($NO_2$) emissions, which are also mainly produced by fossil fuel combustion. Tropospheric $NO_2$ column concentration data from satellites[17,18], and surface $NO_2$ concentrations from air quality stations show a decrease (See *Methods*, SI Figures S7 and SI Table 2) consistent with the reduction of fossil carbon fuels emissions presented above. Overall, $NO_2$ decreased over China in the first four months of 2020 is consistent with our calculated NO2 emission declines based on near real time activity and emission data (See Methods and SI Table S9),) . Over the U.K., France, Germany, and Italy, $NO_2$ decreased by a similar amount than in the U.S. Over India, $NO_2$ showed a weaker decline, also consistent with satellite data.

Overall, $NO_2$ declines in January and February over China are also the largest declines since the OMI data become available in 2004 (SI Figure S10). The consistent results from both ground based and satellite monitoring systems confirm the significant decline of the NO2 concentrations due to COVID-19(See Methods). Based on the OMI satellite data, Over the U.K., France, Germany, and Italy, $NO_2$ decreased by a similar amount than in the U.S.. Over India, $NO_2$ showed a weaker decline, also consistent with satellite data.

**Discussion**

It is still unclear to what extent annual $CO_2$ emissions will be continue to be affected by the COVID-19 pandemic, which will depend on the efficacy and stringency of public health policies and the recovery of economies and human activities around the world. The IMF predicts that the global annual economic output (GDP) will decrease by -3.0% in 2020, which is worse than the financial crisis in 2008 [19], and yet this projection was based on the assumption that the COVID-19 epidemics will fade globally in the second half of this year. Based on near real time activity and emission data, we estimate a decrease of 7.8% of global CO2 emissions in first 4 months of 2020, the largest ratio of decrease ever recorded, larger than during the 2009 economic crisis. However, such reduction so far have very limited impacts on global CO2 concentration, with the parts per million (ppm) molecules of CO2 concentration only decreased by 0.13ppm (the current CO2 concentration is 416ppm) given that 1 part per million of atmospheric CO2 is equivalent to 2.13 Gigatonnes Carbon. More importantly, with observed fast recovery of CO2 in China and the reopening of economy globally, the annual decrease rate of CO2 would be expected to less than 8%, yet the long term trends are still unknown.

Nevertheless, given negative impacts on the carbon intensive industry sector such as cement production, we inferred improvements of the ratio of the emission intensity (CO2 emissions per unit of GDP) in China (3.5%), US (4.5%) and Europe (1.8%), although such improvement are the consequence of highest-ever cost payed for the reduction of 1t of CO2 - between 1k and 10k US$ per ton, it still suggesting a unique opportunity for green investments and low carbon development in the years to come, for which global concerted efforts will urgently be needed. The ability to monitor trends in emissions in near real time that we demonstrate here will be invaluable in adaptively managing the transition.

## Methods

### CO$_2$ emission in baseline year 2019

The CO$_2$ emissions and sectoral structure in 2018 for countries and regions are extracted from EDGAR V 5.0[10], and the emissions are scaled to the year 2019 based on the growth rates from Liu et al.[20] and the Global Carbon Budget 2019[21]. For countries with no current estimates of emission growth rates in 2019 such as Russia, Japan and Brazil, we assume their growth rates of emissions were 0.5% based on the emission growth rates of rest of world [21].

Given the large uncertainty of CO$_2$ emission in China[22,23], we calculated CO$_2$ emissions based on the methodology developed[24] previously:

$$\text{Emissions} = \sum\sum\sum (Energy\ consumption\ data_{i,j,k} \times Emission\ factors_{i,j,k}) \quad (1)$$

*i, j, k* reflect the regions, sectors and fuel types respectively. In our calculation, *i* covers countries. *j* covers four sectors that are power generation, industry, transportation and household consumption, *k* covers three primary fossil fuel types which are coal, oil and natural gas. Emission factors can be further separated into the net heating values for each fuel "v", the energy obtained per unit of fuel (TJ per t fuel), the carbon content "c" (t C TJ-1 fuel) and the oxidization rate "o", which is the fraction (in %) of fuel oxidized during combustion and emitted to the atmosphere.

$$Emission = \sum\sum\sum (Energy\ consumption\ data_{i,j,k} \times v_{i,j,k} \times c_{i,j,k} \times o_{i,j,k}) \quad (2)$$

For China, the energy consumption of coal, oil and gas in 2000-2017 are based on energy balance tables from China Energy Statistical Yearbook[2]. However, due to the two years lag of the publications of China Energy Statistical Yearbook, we project the energy consumption of coal, oil and gas in 2018 and 2019 by multiplying the annual growth rates of coal, oil and gas reported on the Statistical Communiqué[3]. Country-specific emission factors are adopted in the calculation, which are relatively lower to IPCC default emission factors[6].

We assumed that the emission factors and the structure remain unchanged for each country in 2020 when comparing with 2019. Thus, the rate of change of the emission is calculated based solely on the change of the energy consumption data in 2020 compared to the same period of 2019.

Based on the assumption of sectoral carbon intensity and energy structure remain unchanged from 2018, the EDGAR sectors were aggregated into several main sectors, including power sector, ground transport sector, industry sector, residential sector, aviation sector and international shipping sector.

### Power sector.

For China, we used daily thermal generation data in China to calculate the daily emission changes. For India, daily total electricity generation data by production types are updated by Power System Operation Corporation Limited (https://posoco.in/reports/daily-reports/), and we calculated the thermal production by aggregating the electricity production by *Coal*, *Lignite*, and *Gas, Naphtha & Diesel*. For US, we used daily thermal generation data of 48 states from Energy Information Administration's (EIA) Hourly Electric Grid Monitor (https://www.eia.gov/beta/electricity/gridmonitor/). For EU countries and UK, electricity generation data by production types are collected from ENTSO-E Transparent platform (https://transparency.entsoe.eu/dashboard/show). Due to the poor data quality or missing data, Croatia, Cyprus, Ireland, Luxembourg and Malta are excluded from the calculation. The hourly data of the other 23 EU countries (including Austria, Belgium, Bulgaria, Czech Republic, Denmark, Estonia, Finland, France, Germany, Greece, Hungary, Italy, Latvia, Lithuania, Netherlands, Poland, Portugal, Romania, Slovakia,

Slovenia, Spain, Sweden) and United Kingdom are aggregated into daily electricity generation data. The data cleaning and aggregation method: Electricity generation data can be divided into 3 categories in terms of the time interval of data sampling: 15 minutes, 30 minutes and 60 minutes, divided into 2 categories in terms of data availability: missing values (represented by n/e in the entire database), not-a-number values (represented by N/A and void in the entire database). For duplicate data in the same time, the average value for each power sectors is used for further calculation. Prior to aggregation, data cleaning is conducted by preserving missing values for further per-sector analysis and detecting anomaly values in the data for every time interval of one day (24 hours) and for each power sector, which is named as daily power matrix. For each daily power matrix, the anomaly detection algorithm first assumes all not-a-number values as 0, subsequently uses a detection criteria called modified MAD (Median Absolute Deviation):

$$mMAD = \frac{-1}{\sqrt{2}erfcinv(\frac{3}{2})} median(A - median(A)) \quad (3)$$

For each power sector, the data is consisted of one column vector A, if one element in A exceeds median number of A by 3 times MAD (for float precision it is minused 0.000001), it is regarded as anomaly numbers, deleted in A and replaced by the mean value of other elements. Finally in the aggregation, the mean value of the newly-created A represents the average power, which is multiplied by daily sampling times to get daily electricity generation. The erfcinv function is provided by Python Scipy package. Data-related Procedures are programmed in Python. For Russia, daily electricity generation data are collected from United Power System of Russia (http://www.so-ups.ru/index.php). For Japan, we collect the daily electricity generation data by summarizing electricity data from 10 electricity providers in Japan (Hokkaido Electric Power, Tohoku Electric Power Network, Tokyo Electric Power Company, Chubu Electric Power Grid, Hokuriku Electric Power Transmission & Distribution Company, Kansai Electric Power, Chugoku Electric Power Company, Shikoku Electric Power Company, Kyushu Electric Power and Okinawa Electric Power Company). For Brazil, daily electricity generation data by production types are downloaded from the Operator of the National Electricity System (http://www.ons.org.br/Paginas/). We calculate the national emission changes based on the changes of daily thermal production, or total electricity generation data when thermal production data are not available (i.e. Russia and Japan).

For countries not listed above, we estimate the emission changes based on the start time of the national closures. We classify the countries in rest of world according to whether they adopt closure measures. The national closure policies data are collected from GCP, Wikipedia and related News. Based on the daily emission data of power and industry sectors of the countries estimated in this study, we calculate the change rates of electric power and industrial sectors in the countries with closures and the countries without closures between January and April respectively. In order to estimate the daily emission changes of rest of world in 2020, we first calculate the average daily emissions based on the national annual emission estimates in 2018 published by EDGAR. For the countries with closures, we use the corresponding emission reduction rates to revise daily estimated emissions during the period of closures. For the countries without closures, we use the corresponding reduction rates to revise. We further aggregate the daily emission changes of rest of world and calculate the total reduction rate of rest of world caused by closures compared with the normal situation. Using this decline rate and daily emission changes, we estimate the daily emission changes of rest of world in 2020 based on the total emission data of rest of world published by GCP in 2019.

**Industry and cement production.**

For China, the industrial sector was divided into four sub-categories including steel industry, cement industry, chemical industry, and other industries, based on the structure of industrial emissions calculations conducted by IEA[25]. For each category, the monthly production data was obtained and the corresponding month-on-month growth rate was calculated accordingly. Specifically, the production data was regarded as the emission estimator i.e. activity while the emission factors were assumed the same as 2019 for each industry. For the steel industry, we collected the global monthly crude steel production data from World Steel Association website (https://www.worldsteel.org/) while the monthly production data of cement, chemicals as well as other industries were referred to the National Bureau of Statistics website. For the latter two multi-component categories, the production of sulfuric acid, caustic soda, soda ash, ethylene, chemical fertilizer, chemical pesticide, primary plastic and synthetic rubber was taken into account while 26 other industrial products including crude iron ore, phosphate ore, salt, feed, refined edible vegetable oil, fresh and frozen meat, milk products, liquor, soft drinks, wine, beer, tobaccos, yarn, cloth, silk and woven fabric, machine-made paper and paperboards, plain glass, ten kinds of nonferrous metals, refined copper, lead, zinc, electrolyzed aluminum, industrial boilers, metal smelting equipment, and cement equipment were included in the other industries sub-group. The calculation of growth rate for the steel and cement industries was relatively straightforward, in that the 2020 and 2019 month-on-month data were compared. In terms of the latter two multi-component groups, the growth rates were evaluated based on the weighed contribution from each product. Based on the emission distribution of these four industries in the industrial sector in China in year 2019 and the growth rates obtained as stated above, we finally estimated the monthly emission in the first four months of 2020.

For US, Europe (EU27 & UK), Japan, Russia, India and Brazil, we use the cumulative Industrial Production Index (IPI) to estimate the growth rates of emissions in these countries or regions, collected from Federal Reserve Board (https://www.federalreserve.gov), Eurostat (https://ec.europa.eu/eurostat/home), Ministry of Economy, Trade and Industry (https://www.meti.go.jp), Federal State Statistics Service (https://eng.gks.ru), Ministry of Statistics and Programme Implementation(http://www.mospi.nic.in) and Brazilian Institute of Geography and Statistics (https://www.ibge.gov.br/en/institutional/the-ibge.htm) respectively. However, the last observation in Europe, Japan, Russia, India and Brazil were in March 2020. To estimate the current growth in April 2020, for Japan, Russia and Brazil, we adopt the predicted results from Trading Economics(https://tradingeconomics.com). For Europe and India, we use the average of Japan, Russia and Brazil's growth rate in April 2020 to estimate their current growth because of the implement of locking down polices similar to those three countries. Based on the growth rates, we calculate the monthly data of industrial sector for 2019 and the first four months of 2020. The monthly industrial emissions are allocated to daily emissions by daily electricity data. We follow same measure for power generation to calculated the emission from industry and cement production for rest of the world.

**Ground transportation**.

We collected TomTom congestion global level data from TomTom website (https://www.tomtom.com/en_gb/traffic-index/). The congestion level (called X hereafter) represents the extra time spent on a trip, in percentage, compared to uncongested condition. TomTom congestion level data are available for 416 cities across 57 countries at a temporal resolution of one hour to 15 minutes. Of note that a zero congestion level means that the traffic is fluent, but rather than no cars and zero emissions. It is thus important to identify the low threshold of emissions when the congestion level is zero. We compared the time series of daily mean TomTom congestion level with the daily mean car counts (called Q hereafter) on main roads in Paris. The daily mean car counts were reported by the City's service (https://opendata.paris.fr/pages/home/). We used a sigmoid function to describe the relationship between X and Q (SI Fig S5):

$$Q = a + (bX^c)/(d^c + X^c) \quad (4)$$

where a, b, c and d are the regression parameters. It is shown that the regression can reflect large drop down in the ground transportation due to the lockdown and the recovery afterwards. We assume that the daily emissions were proportional to this relative magnitude of daily mean car counts. Then, we applied the regression built for Paris to other cities included in TomTom dataset, assuming that the relative magnitude in car counts (and thus emissions) follow the similar relationship with TomTom. We compared the time series of TomTom congestion level in the first quarter of 2019 and 2020. The emission changes were first calculated for individual cities, and then weighted by city emissions to aggregate to national changes. The weighting emissions are taken from the gridded EDGARv4.3.2 emission map for the "road transportation" sector (1A3b) (https://edgar.jrc.ec.europa.eu/) for the year 2010, assuming that the spatial distribution of ground transport do not change significantly within a country. For countries not included in the TomTom dataset, we assume that the emission changes follow the mean changes of other countries. For example, Cyprus, as an EU member country, is not reported in TomTom dataset, and its relative emission change was assumed to follow the same pattern of the total emissions from other EU countries included in TomTom dataset (which covers 98% of EU total emissions). Similarly, the relative emission changes of countries in ROW but were not reported by TomTom were assumed to follow the same pattern of the total emissions from all TomTom countries (which cover 85% of global total emissions). The uncertainty in the TomTom-based Q, and thus emissions were quantified by the prediction interval of the regression.

**Aviation.**

Although both the International Civil Aviation Organization (ICAO) and the International Air Transport Association (IATA) publish yearly statistics of aircraft operations, CO2 emissions from commercial aviation are usually reconstructed from bottom up emission inventories based on the knowledge of the parameters of individual flights. The International Council on Clean Transportation (ICCT) published that CO2 emissions from commercial freight and passenger aviation resulted in 918 Mt CO2 in 2018 (Graver et al., 2019) based on the OAG flight database and emission factors from the PIANO database. IATA estimated a 3.4% increase between 2018 and 2019 in terms of available seat kilometers (see https://www.iata.org/en/pressroom/pr/2020-02-06-01/). In the absence of further information, we consider this increase to be representative of freight aviation as well and use a slightly smaller growth rate of 3% for CO2 emissions between 2018 and 2019 to account for a small increase in fuel efficiency.

Individual commercial flights are tracked by Flightradar24 based on reception of ADS-B signals emitted by aircraft and received by their network of ADS-B receptors; they are provided to us for 2019 and 2020 with monthly updates. As we do not have yet the capability to convert the FlightRadar24 database into CO2 emissions on a flight-by-flight basis, we compute CO2 emissions by assuming a constant CO2 emission factor per km flown across the whole fleet of aircraft (regional, narrowbody passenger, widebody passenger and freight operations). This assumption is justified if the mix of flights between these categories has not changed substantially between 2019 and 2020. The kilometers flown are computed assuming great circle distance between the take-off, cruising, descent and landing points for each flight and are cumulated over all flights. The FlightRadar24 database has incomplete data for some flights and may miss altogether a small fraction of actual flights, so we scale the ICCT estimate of CO2 emissions (inflated by 3% for the year 2019) with the total estimated number of kilometers flown for 2019 (67.91 109 km) and apply this scaling factor to 2020 data. Again this assumes that the fraction of missed flights is the same in 2019 and 2020, which seems reasonable. As the departure and landing airports are known for each flight, we can classify the km flown (and hence the CO2 emissions) per country, and for each country between domestic or international traffic.

**Ships**

According to the Third IMO GHG Study29, CO2 emissions from international shipping accounts for 87% of global shipping emissions, domestic and fishing accounts for 9% and 4%, respectively. We estimated global CO2 shipping emissions from 2016-2018 with the EDGAR's international emissions and the ratio between international shipping and global shipping emissions. And we extrapolated emissions from 2007-2018 to estimate emissions in 2019. We obtained emissions for the first quarter of 2019 based on the assumption that monthly variation is flat in shipping CO2 emissions. In addition, we assume that the change in shipping emissions is linearly related to the change in ship's volume. We estimated the change of shipment by the end of Apr by -15% compared to the same period of last year according to the report (https://www.theedgemarkets.com/article/global-container-shipments-set-fall-30-next-few-months).

### Residential and commercial buildings

The calculation of emissions was performed in three steps: 1) Calculation of population-weighted heating degree days for each country and for each day based on the ERA5 reanalysis of 2-meters air temperature, 2) Using the EDGAR estimates of 2018 residential emissions as the baseline. For each country, the residential emissions were split into two parts, i.e., cooking emissions and heating emissions, according to the EDGAR guidelines. The emissions from cooking were assumed to remain stable, while the emissions from heating were assumed to depend on and vary by the heating demand. 3)

Based on the change of population-weighted heating degree days in each country, we scaled the EDGAR 2018 residential emissions to 2019 and 2020. Since the index of heating degree days are daily values, we can get daily emissions update for the residential sources globally. Note that the effect of increased time spent in households on residential buildings and decreased time in commercial and public buildings was not accounted for, since we did not have fuel consumption data for urban areas and building types. Our estimates of residential emissions changes are consistent with those obtained from the City of Paris, based on individual electricity use (https://data.enedis.fr/) and population surveys (Y. Françoise pers. comm.).

### Disaggregation of the subset of data available on a monthly basis into daily variations

For industrial sector, we calculate the monthly changes based on the monthly statistics, and disaggregate the monthly industrial emissions into daily industrial emissions by daily electricity generation on national level.

### Uncertainty estimates

We followed the 2006 IPCC Guidelines for National Greenhouse Gas Inventories to conduct the uncertainty analysis of the data. Firstly, the uncertainties were calculated for each sector:

Power sector: the uncertainty is mainly from inter-annual variability of coal emission factors. Based the UN statistics the inter-annual variability of fossil fuel is within (±1.5%), which been used as uncertainty of the CO2 from power sectors.

Industrial sector: Uncertainty of CO2 from Industry and cement production comes from the monthly production data. Given CO2 from Industry and cement production in China accounts for more than 60% of world total industrial CO2, and the fact that uncertainty of emission in China is t Uncertainty from monthly statistics was derived from 10000 Monte Carlo simulations to estimate a 68% confidence interval (1-sigma) for China. from monthly statistics was derived from 10000 Monte Carlo simulations to estimate a 68% confidence interval (1-sigma) for China. We calculated the 68% prediction interval of linear regression models between emissions estimated from monthly statistics and official emissions obtained from annual statistics at the end of each year, to deduce the one-sigma uncertainty involved when using monthly data to represent the whole year's change. The squared correlation coefficients are within the range of 0.88 (e.g., coal production) and 0.98 (e.g., energy import and export data), which represent that only using the monthly data can explain 88% to 98% of the whole year's variation34, while the remaining variation not covered yet reflect the uncertainty caused by the frequent revisions of China's statistical data after they are first published.

Ground Transportation: The emissions in ground transportation sector is estimated by assuming that the relative magnitude in car counts (and thus emissions) follow the similar relationship with TomTom. So the emissions were quantified by the prediction interval of the regression.

Aviation: The uncertainty of aviation sector comes from the difference of daily emission data estimated based on the two methods. We calculate the average difference between the daily emission results estimated based on the flight route distance and the number of flights, and then divide the average difference by the average of the daily emissions estimated by the two methods to obtain the uncertainty of CO2 from aviation sector.

Shipping: We used the uncertainty analysis from IMO as our uncertainty estimate for shipping emissions. According to Third IMO Greenhouse Gas study 201431, the uncertainty of shipping emissions was 13% based on bottom-up estimates.

Residential: The 2-sigma uncertainty in daily emissions are estimated as 40%, which is calculated based on the comparison with daily residential emissions derived from real fuel consumptions in several European countries including France, Great Britain, Italy, Belgium, and Spain.

The uncertainty of emission projection in 2019 is estimated as 2.2%, by combining the reported uncertainty of the projected growth rates and the EDGAR estimates in 2018.

Then we combine all the uncertainties by following the error propagation equation from IPCC. Equation 5 is used to derive for the uncertainty of the sum, which could be used to combine the uncertainties of all sectors:

$$U_{total} = \frac{\sqrt{\sum(U_s \cdot \mu_s)}}{|\sum \mu_s|} \qquad (5)$$

Where $U_s$ and $\mu_s$ are the percentage uncertainties and the uncertain quantities (daily mean emissions) of sector $s$ respectively.

Equation 6 is used to derive for the uncertainty of the multiplication, which is used to combine the uncertainties of all sectors and of the projected emissions in 2019:

$$U_{overall} = \sqrt{\sum U_i^2} \qquad (6)$$

Table 6 **Percentage uncertainties of all items.**

| Items | Uncertainty Range |
|---|---|
| Power | ±1.5% |
| Ground Transport | ±9.3% |
| Industry | ±36.0% |
| Residential | ±40.0% |
| Aviation | ±10.2% |
| International Shipping | ±13.0% |
| Projection of emission growth rate in 2019 | ±0.8% |
| EDGAR emissions in 2018 | ±5.0% |
| **Overall** | **±6.8%** |

**Satellite observation and data sources:**

To validate the response of the atmosphere, including $CO_2$ concentration and air quality, to the decreased fossil fuel burning and transportation, we collected $NO_2$, aerosol optical depth (AOD) and column-averaged dry air mole fraction of $CO_2$ ($XCO_2$) data from satellites ($NO_2$ from OMI, AOD from MODIS and $XCO_2$ from GOSAT) and surface daily average nitrogen dioxide ($NO_2$, µg/m$^3$), carbon monoxide (CO, µg/m$^3$) from 1600 air quality monitoring sites (China and US, SI Figure S7) in to investigate the impact of COVID-19 on air quality and atmospheric $CO_2$.

Surface air quality data in China was collected from the daily report by Ministry of Ecology and Environment of China (http://www.mee.gov.cn/). Measurements of daily average nitrogen dioxide (NO2, µg/m3), carbon monoxide (CO, µg/m3), and particulate matter smaller than 2.5 µm (PM2.5, µg/m3) from 1580 sites used to estimate pollution changes between the first quarters of 2019 and 2020. Surface air quality data in U.S. is downloaded from the Air Quality System (AQS) operated by the U.S. Environmental Protection Agency (https://www.epa.gov/aqs). Measurements of daily maximum 1-hour NO2 (ppb), daily maximum 8-hour CO (ppm), and daily average PM2.5 (µg/m3) from 983 sites are used. For March 2020, data availability is limited in U.S. with 20 sites for NO2, 31 for CO, and 309 for PM2.5. Sites with missing data for NO2/CO (PM2.5) at over 20 (5) days in any months will be excluded.

We obtained monthly $NO_2$ data from the Ozone Monitoring Instrument (OMI) provided by Tropospheric Emission Monitoring Internet Service (TEMIS), which has with a spatial resolution of 0.125° x 0.125° and a temporal coverage from October 2004 to March 2020. We only included the data from January 2013 to March 2020 in the work (SI Figure S8). For AOD, we chose daily Level 2 MOD 04 data from MODIS[34] and then calculated the monthly averaged AOD f from January 2013 to March 2020. Only "good" and "very good" data (in AOD_550_Dark_Target_Deep_Blue_Combined_QA_Flag 2 and 3) were kept in the calculation. At last, we calculated the monthly $XCO_2$ data with a resolution of 2.5° x 2.5° from the Greenhouse Gases Observing Satellite "IBUKI" (GOSAT). Because of the delay in the data processing at National Institute for Environmental Studies (NIES), we used a bias-uncorrected version V02.81 for the period of January 2013 to March 2020. With the consideration of the focus on an abnormal event due to COVID-19, the bias-uncorrected data is proper for this study.

All of the monthly averaged data were re-gridded to 1° x 1°. We focused on four emitting regions, China, USA, EU4 (UK, France, Germany, and Italy), and Indian, and then calculated the country level monthly averaged $NO_2$, AOD, and $XCO_2$ values.

Surface air pollution in China was significantly reduced during the epidemic period (SI Figure 7). A deep reduction of NO2 by 31.7% was observed on January 24th 2020, one day after the lockdown for many provinces (Si Figure S7b). The reduction rates were 13.7% for PM2.5 and 16.5% for CO on the same day. A clear rebound (U shape) could be found for all pollution after the spring festival (February 5th) in 2019. However, such recovery was missing in 2020 due to the lockdown, leading to a decreasing trend all through the first quarter. On average, pollution concentrations decreased by 23.0% for NO2, 15.4% for PM2.5, and 12.5% for CO during January-March 2020 relative to the same period in 2019.

Pollution level in U.S. was also reduced by the epidemic but with smaller magnitude compared to that in China. Surface PM2.5 decreased in all first three months in 2020 relative to 2019 with the largest reduction of 20.6% in March. NO2 also exhibited large reductions of 9.0% on March 2020 compared to 2019, however, such reduction seemed affected by the limited site numbers (only 20). For example, one site in Salt Lake, Utah reported >200 ppb (normally <40) NO2 during March 20-23, 2020. Such episodes were likely caused by fires but weakened the reduction rate of NO2 after Middle March (SI Figure 7e). Changes of CO were also limited in U.S., with opposite signs in January and March. Such tendencies could also be biased due to the limited site numbers (only 31).

The observed tropospheric nitrogen dioxide ($NO_2$) column concentration data from satellite observation[35] and surface air quality data from ground monitoring networks have exhibited a decrease (SI Table S9) consistent with reduction of fossil carbon fuels emissions.

In China, January, February, March, and April 2020 decreased by -32.27%, -34.22%, -4.53%, -3.59% respectively compared to 2019. Overall, NO2 decreased over China by -21.55% from January to April 2020 compared to 2019. In the US, the decrease of $NO_2$ first started in Feb and continued to decrease at least until March 2020. Compared to the same period of the year in 2019, $NO_2$ over the US decreased by -23.08% and - 14.32% in February and March 2020, respectively (SI Table S9). For the UK, France, Germany, and Italy, we observe similar $NO_2$ decreases than over the USA India had weaker decline in $NO_2$ than other regions.

The decline rate of $NO_2$ (-21.55%) based on atmospheric observations can be used to check the consistency of the decrease of $NO_2$ emission from the inventory, and given the NO2 is mainly contributed by fossil fuel combustion with life time short than one day, the temporal change of NO2 emission can could verify the decrease of the fossil fuel combustion and the associated $CO_2$ emissions. For China where the most significant decrease of tropospheric $NO_2$ column concentration observed, the inventory- based estimates[36,37] of power generation (-6.8%), transportation (-37.2%) and industry(-8.1%) are adopted with result of weight mean -23.94% % NO2 emission in first quarter of 2020 when comparing with 2019. These three sectors together account for 96% of China's total NO2 emissions. The -23.42% decline of the $NO_2$ emissions from our bottom-up inventory is consistent with the satellite observed -26% decrease of column $NO_2$, and with the -23% decrease of near surface concentrations at the 1680 ground-based stations. For US, the inventory-based estimates of power generation (-4.9%), transportation (-2.7%) and industry(-2.2%) are adopted with result of -2.57% NO2 emission in first quarter of 2020 when comparing with 2019, slightly smaller than -4.76% tropospheric $NO_2$ column concentration, but difference with the site observation data (-8.98% in March and +0.34% for first quarter), which may be affected by site numbers (only 20 sites in US). We calculated 1° x 1° monthly mean of NO2, AOD, and $XCO_2$ from OMI, MODIS, and GOSAT, respectively (SI Table S9).

For satellite observations, the overall uncertainty of tropospheric NO2 columns monthly mean is 10%[38]. The uncertainty of AOD is approximately $0.03+0.20\tau_M$, where $\tau_M$ is AOD at 550 nm[39]. In other words, the uncertainties in percentage in low AOD regions (US and EU4) is higher in high AOD regions (China and India). The standard deviations of $XCO_2$ monthly mean over land are about 0.5-1.5 ppm[40].
Here we conservatively considered uncertainty of monthly $XCO_2$ as 1.5 ppm. To estimate the uncertainty of changes of 2020 compared to 2019 from January to March, we input above uncertainties of monthly means and run Monte Carlo simulations of 10000 trials to calculate the 68% confidence intervals (i.e., one sigma range) which are shown in Table 1.

**Data Availability Statement**

All data generated or analyzed during this study are included in this article.

**Code Availability**

The code generated during and/or analyzed during the current study are available from the corresponding author. After peer-reviewed the code will be open accessible on the Carbon Monitor website ([www.carbonmonitor.org](www.carbonmonitor.org) or [www.carbonmonitor.org.cn](www.carbonmonitor.org.cn)).

**Completing Interests statement**

Authors declare no competing interests.

**Author contribution:**

Zhu Liu and Philippe Ciais designed the research, Zhu Deng coordinated the data processing. Zhu Liu, Philippe Ciais and Zhu Deng contributed equally in this research, all authors contributed to data collecting, analysis and paper writing.

Authors acknowledge Paul O. Wennberg for insightful comments to improving this paper.

**Supplementary Information**

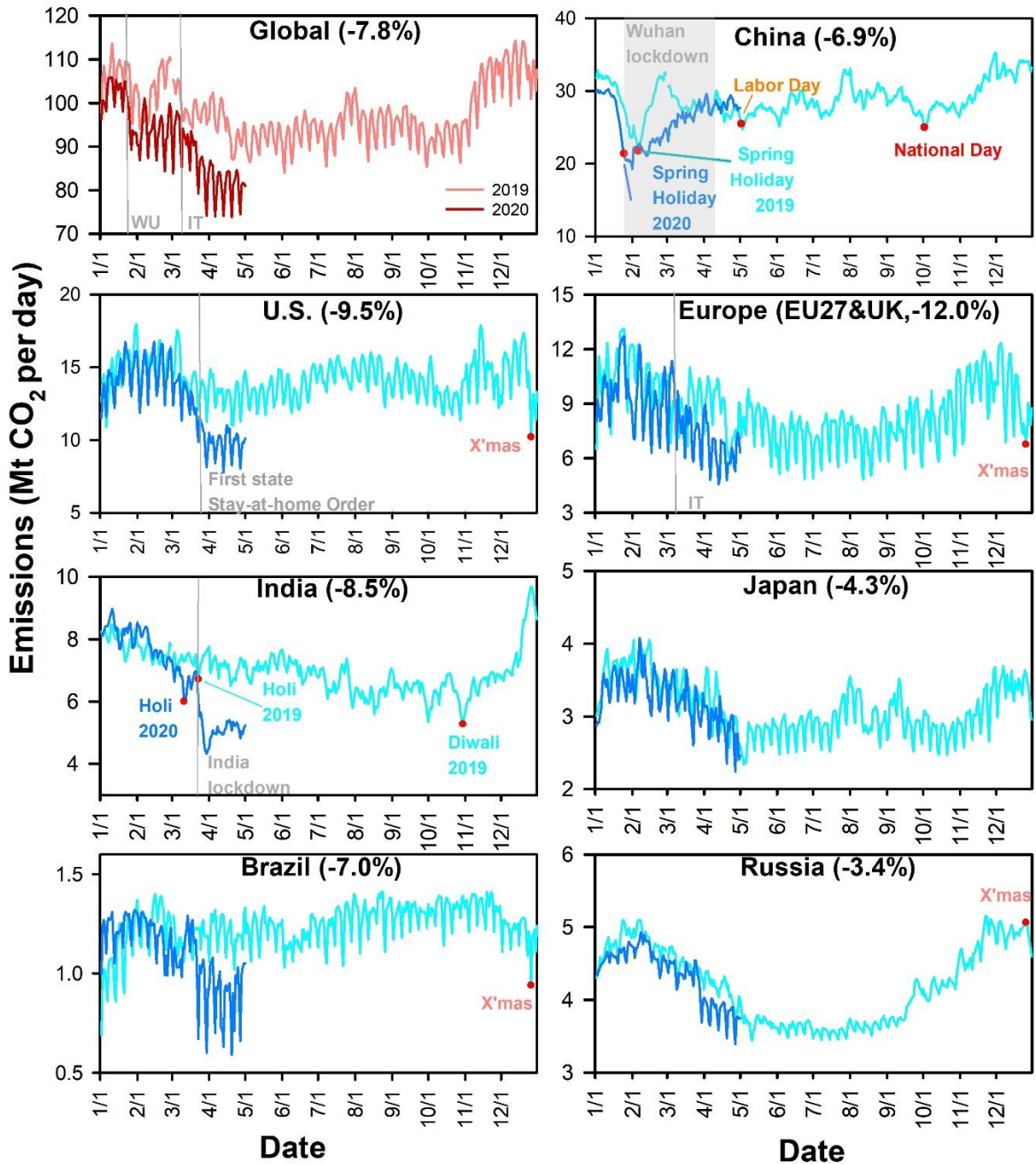

**SI Figure S1. Daily CO$_2$ emissions in 2019 and the first four months of 2020 by countries.**

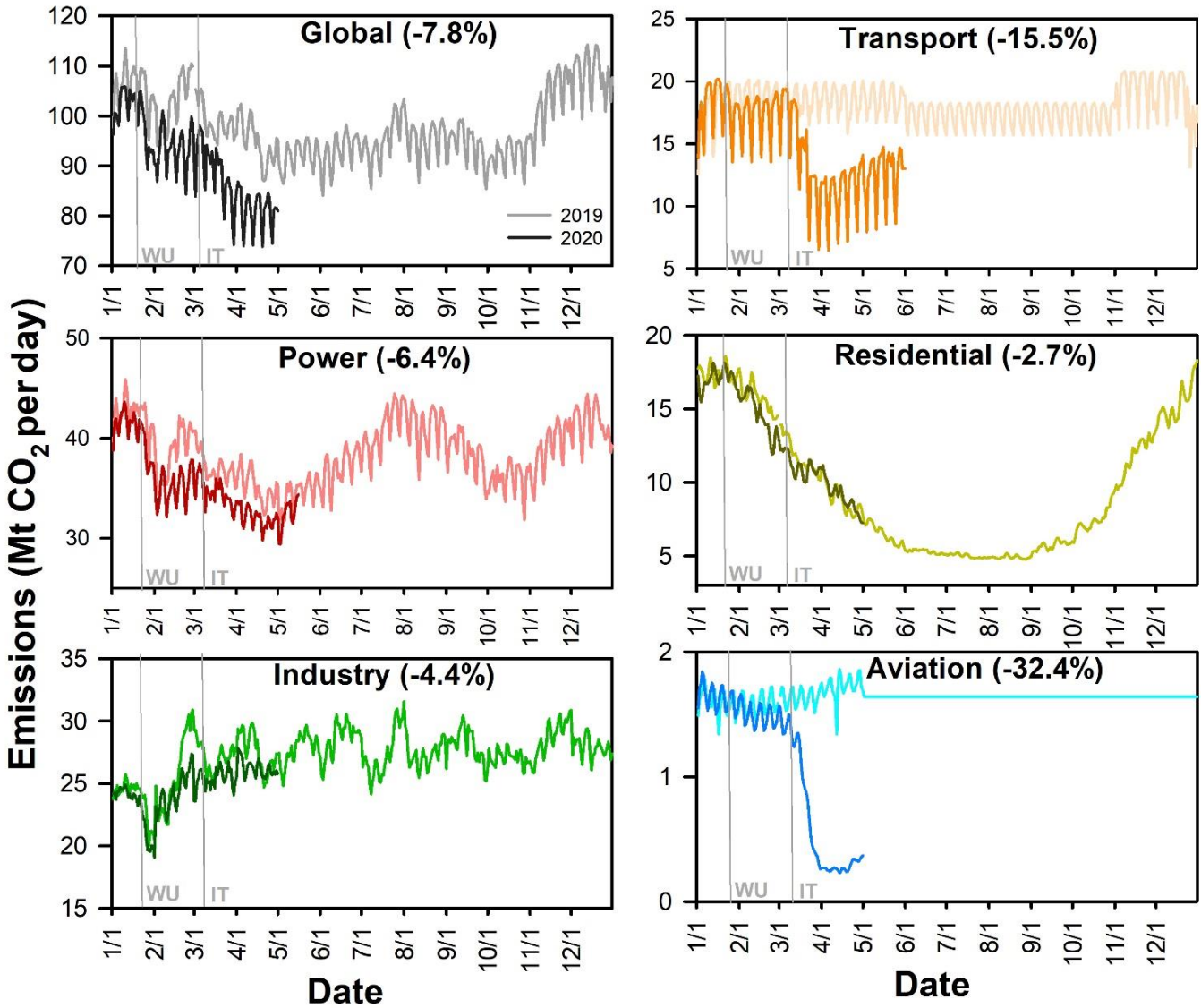

**SI Figure S2 Daily CO2 emissions in 2019 and the first four months of 2020 by countries**

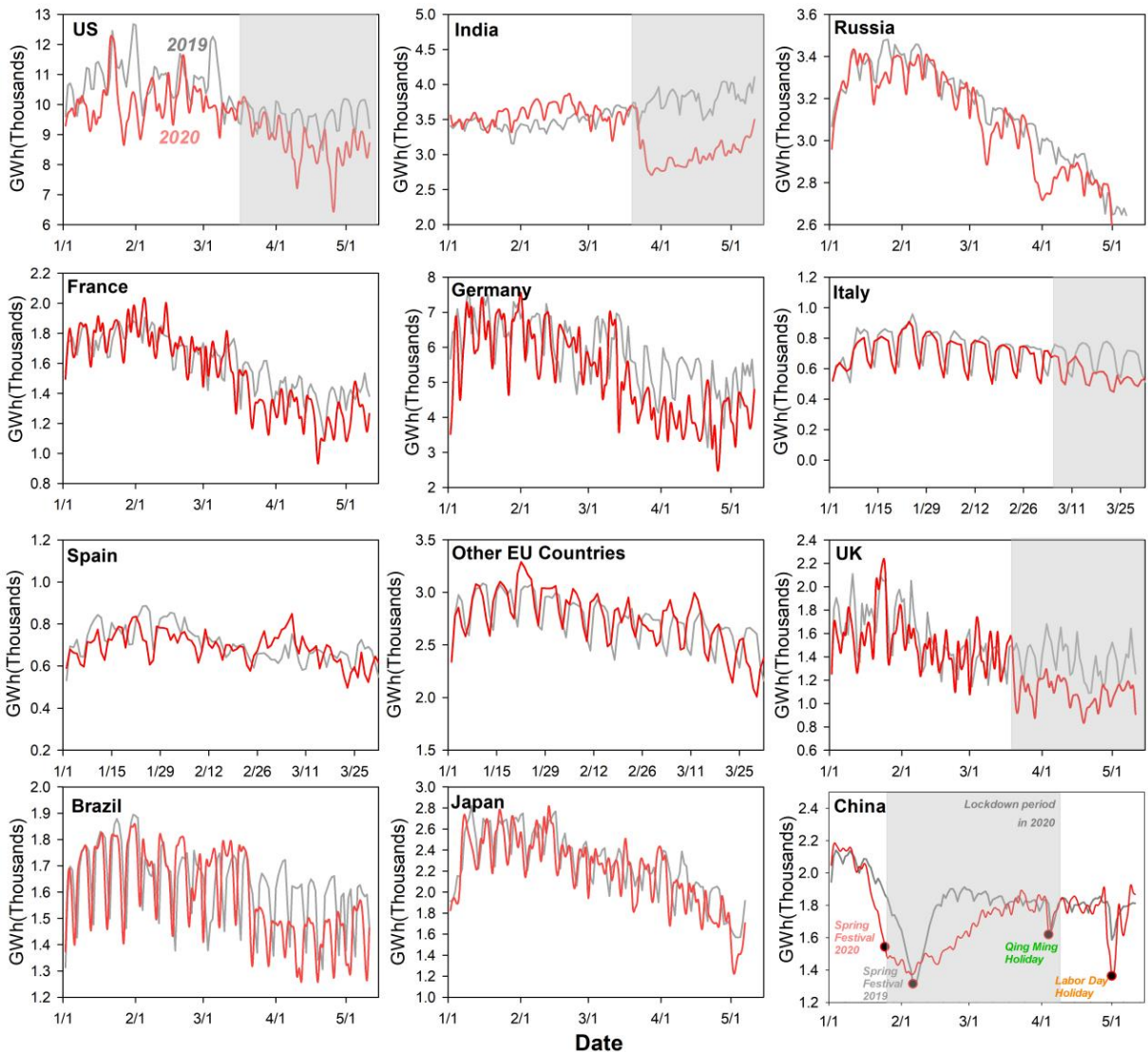

**SI Figure S3 | Daily electricity generation in 2020 in US, India, Russia, France, Germany, Italy, Spain, other European countries (Austria, Belgium, Bulgaria, Czech Republic, Denmark, Estonia, Finland, Greece, Hungary, Latvia, Lithuania, Netherlands, Poland, Portugal, Romania, Slovakia, Slovenia, and Sweden), UK, Brazil, Japan and China, after being corrected by daily temperature (See Methods).**

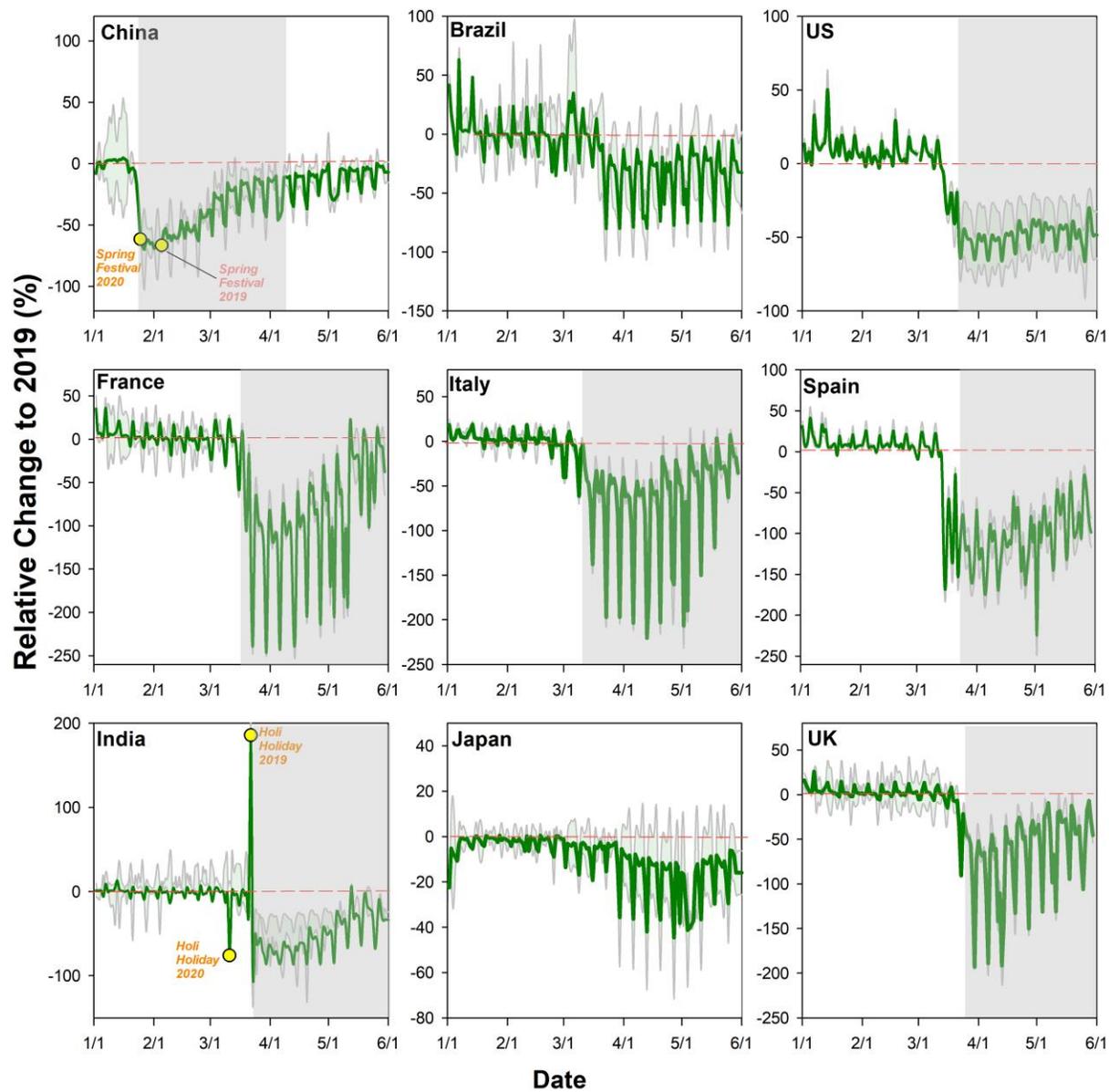

**SI Figure S4 | Monthly emission changes in transport sector in February, March and the first quarter of 2020. (See SI Table 3)**

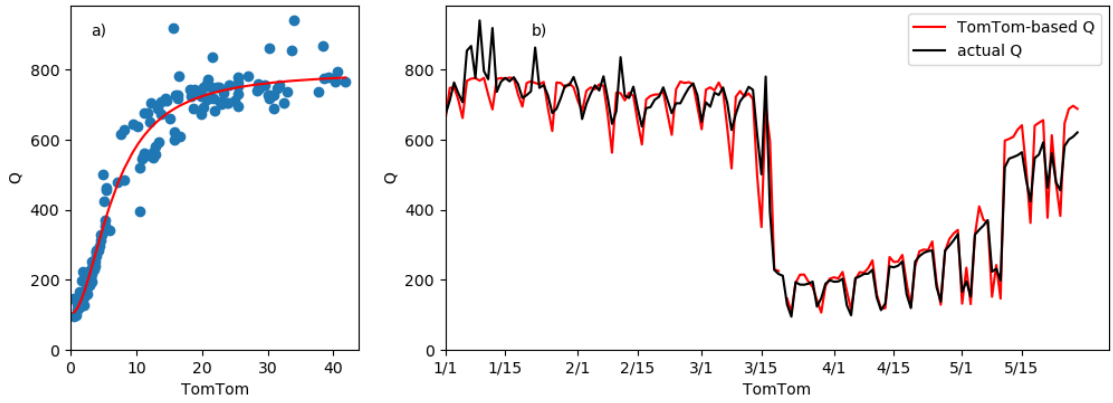

**SI Figure S6.** The relationship between TomTom congestion level and the actual car counts (Q) for Paris. a) the regression between TomTom congestion level (x-axis) and the car counts (y-axis); b) the Q calculated based on TomTom congestion level (red) and the actual Q.

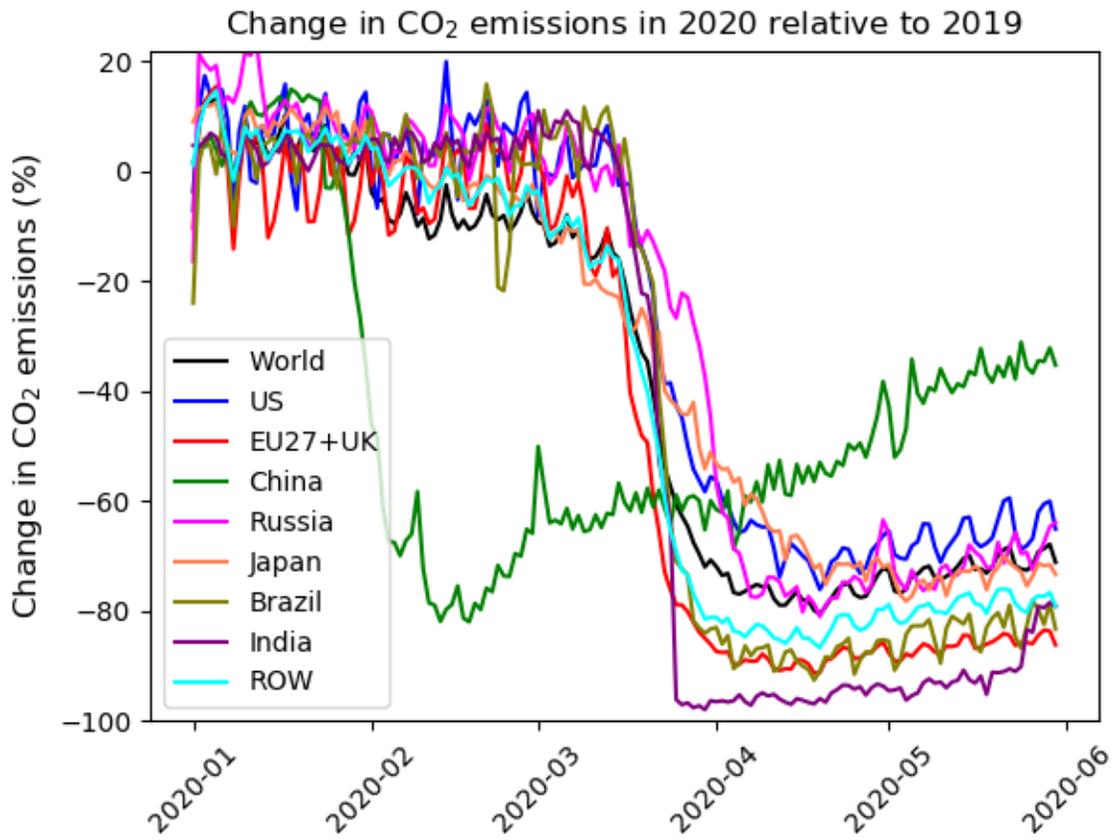

**SI Figure S6. Change in CO2 emissions for aviation.**

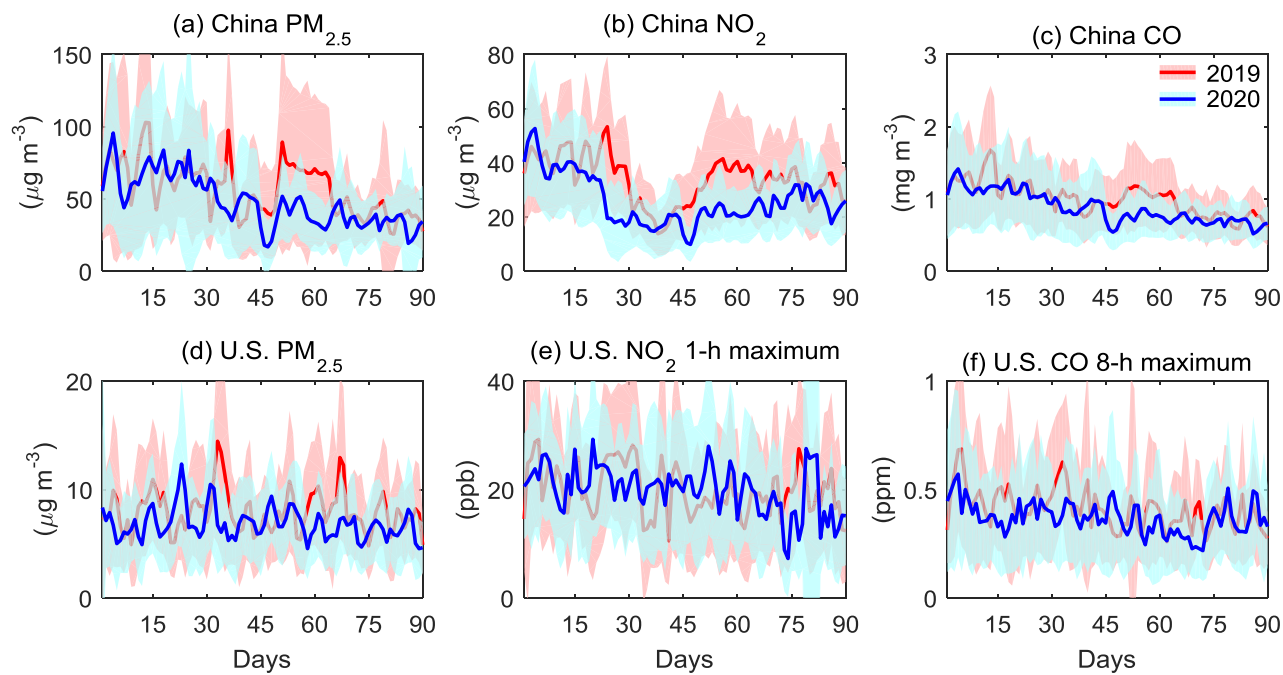

**SI Figure S7.** Daily variations of surface (a, d) PM$_{2.5}$, (b, d) NO$_2$, (c, f) CO concentrations from (a-c) China and (d-f) U.S. during the first quarters of 2019 and 2020. The bold lines are the mean values from all quality-controlled sites, with shadings indicating one standard deviation. The data on February 29$^{th}$ 2020 are removed from the plot.

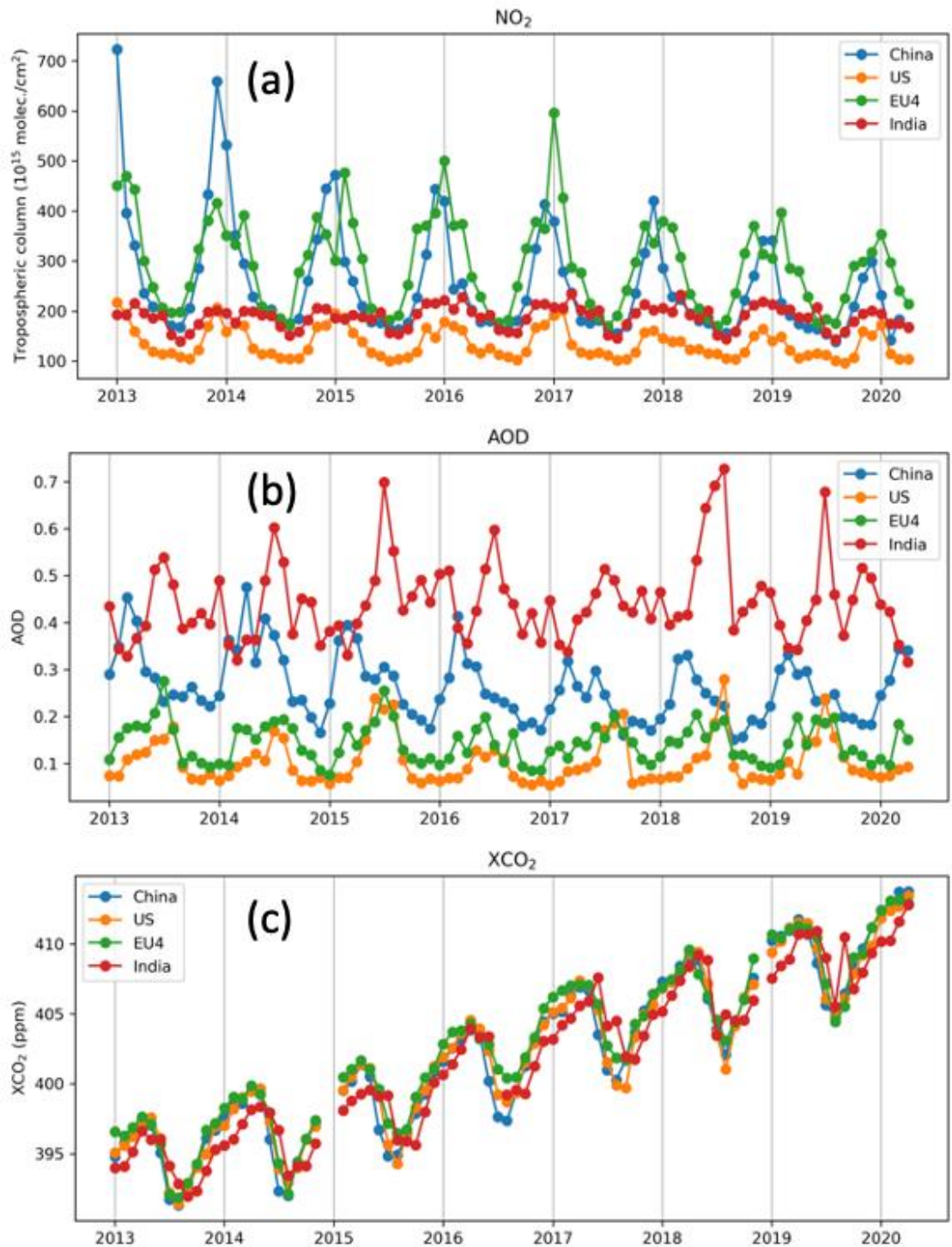

**SI Figure S8**. The monthly series of a) NO2, b) AOD and c) XCO2 over China, US, EU4(UK, Germany, Italy and France), and India

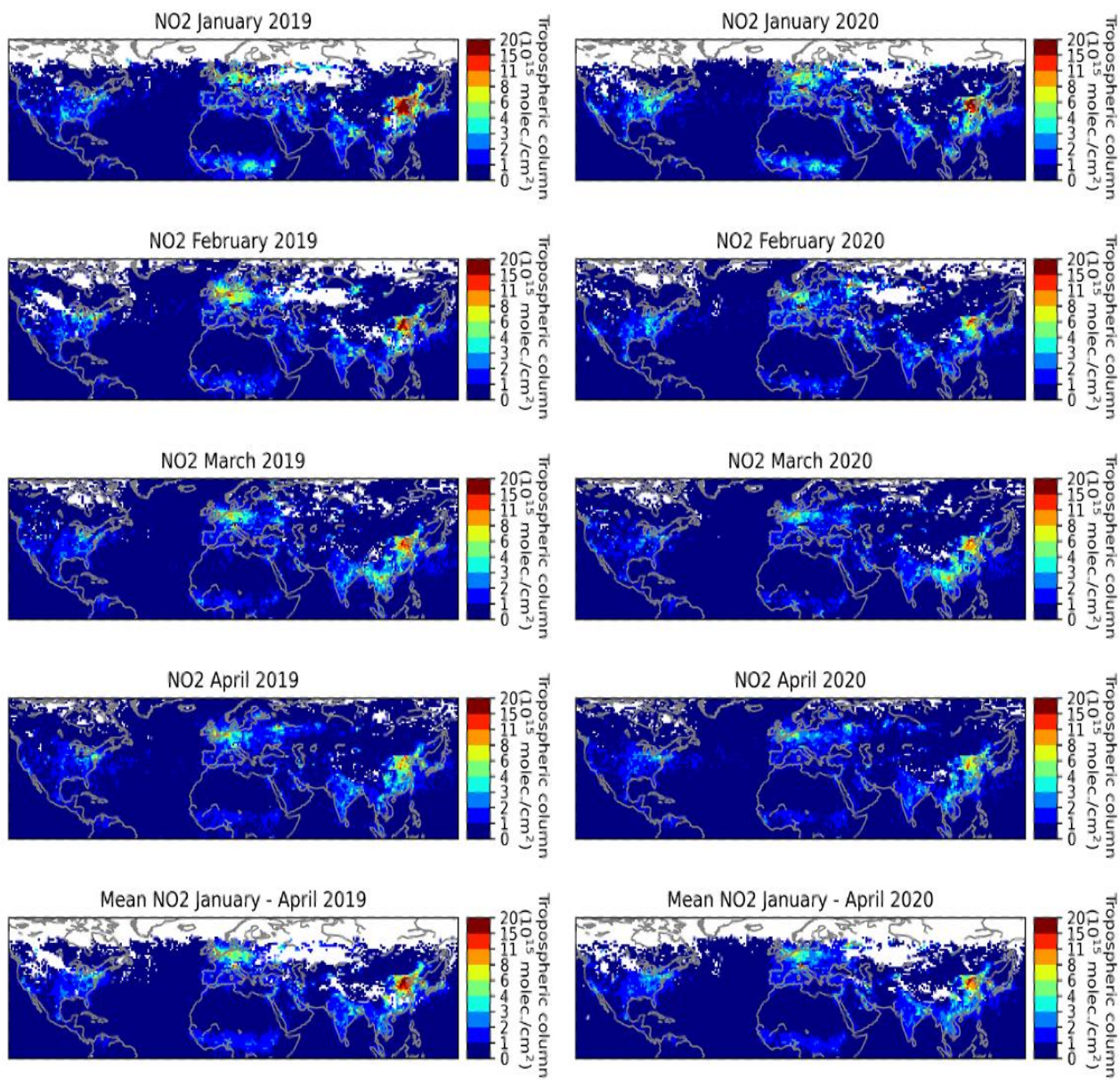

**SI Figure S9 Tropospheric column NO2 Observation in January – April of 2020**

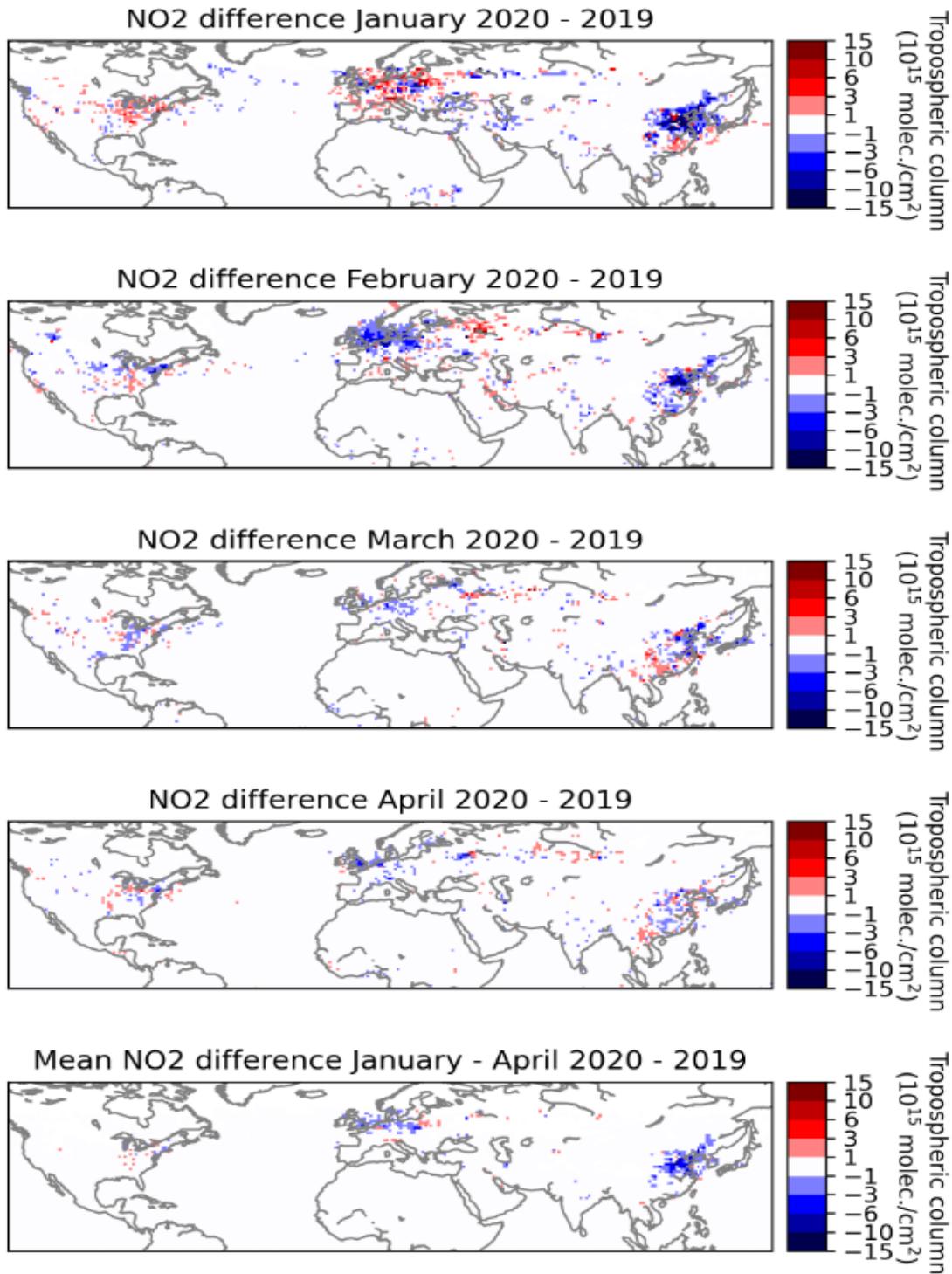

**SI Figure S10 | Tropospheric column NO$_2$ Observation in January – April of 2020**

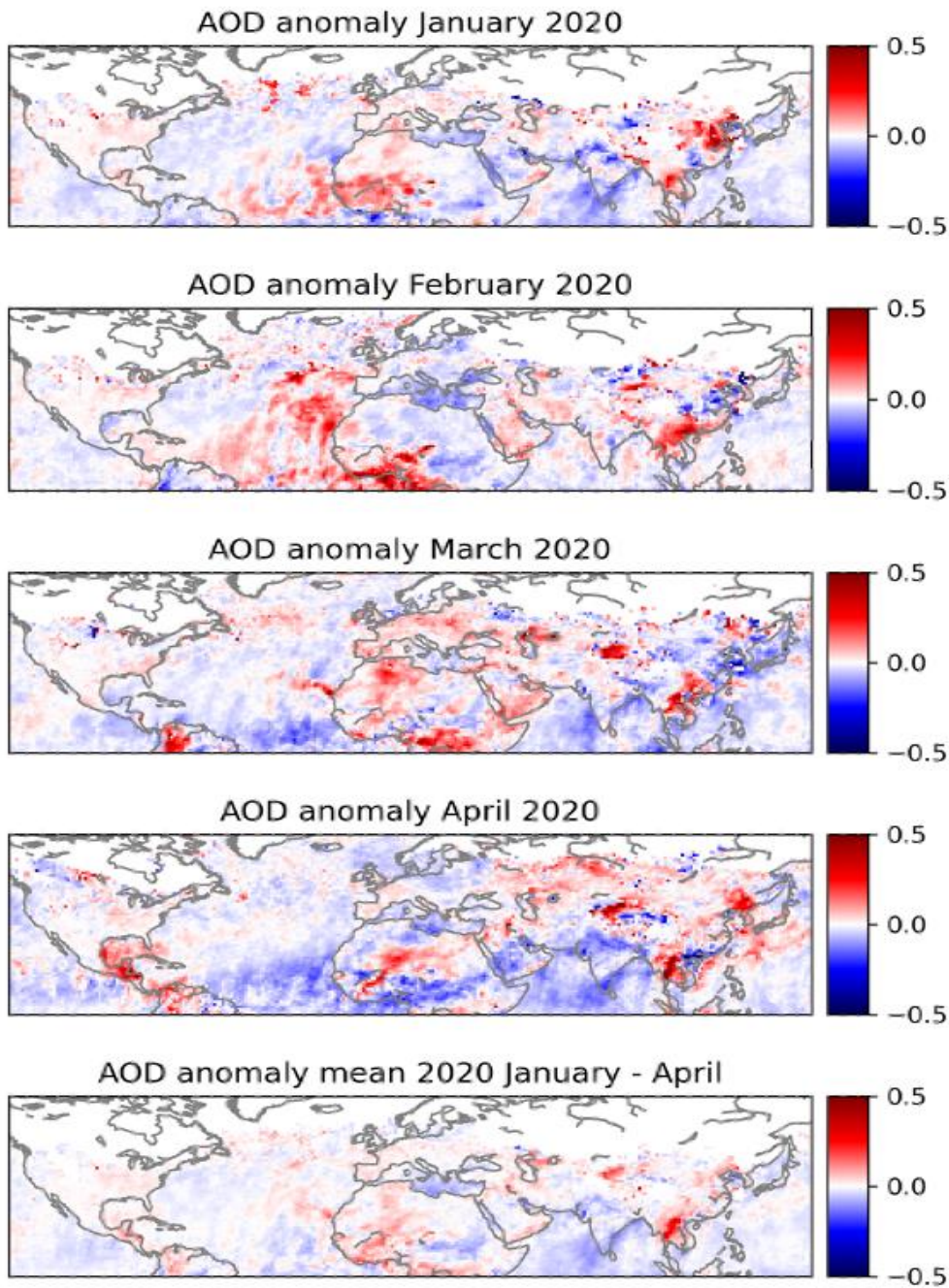

**SI Figure S11. Anomaly of NO2 from OMI in the first quarter of 2020**

The anomaly maps conducted by apply the same algorithm on every grid point. The anomaly defined as the deseasonalized value. For NO2 (Figure S11), the anomaly along the eastern coast of China was negative in January and February 2020, then partially become positive. About half of the anomalies over U.S. and Europe were positive in January 2020, then most areas over U.S. and Europe became negative, which also matches the COVID-19 epidemic delays compared to China.

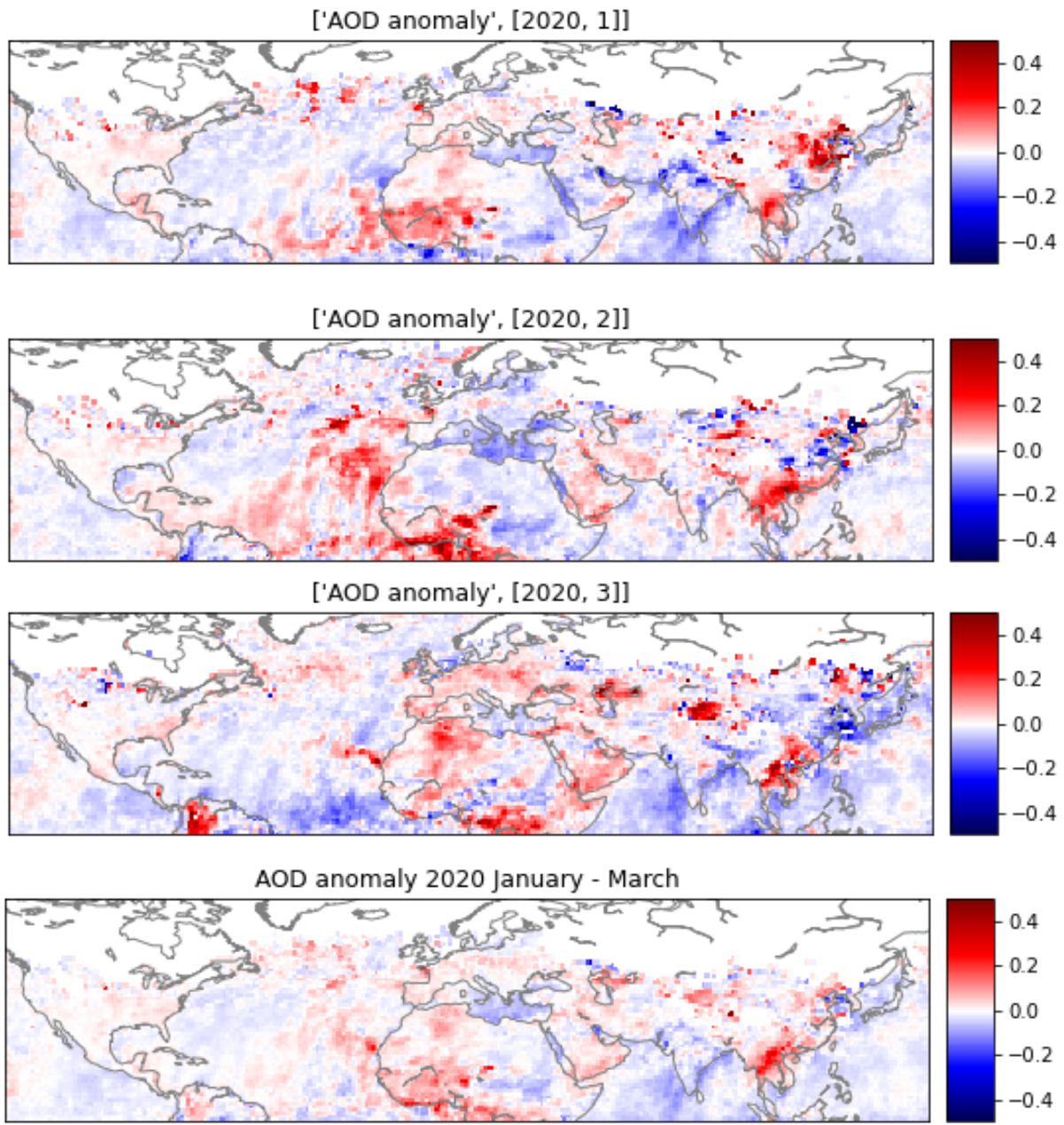

**SI Figure S12 Anomaly of AOD from MODIS in the first quarter of 2020**

The anomaly maps conducted by apply the same algorithm on every grid point. The anomaly defined as the deseasonalized value. For AOD (Figure S12), the negative anomaly area along the eastern coast of China expanded from January to March 2020. For US and Europe, AOD anomalies on land did not change too much. The shutdown of COVID-19 may not affect AOD over them since their AOD was always Low.

**SI Table S1** Mapping table of sectors between this study and EDGAR[9].

| This Study | EDGAR |
|---|---|
| Power | Electricity and heat production |
| Industry (from direct fuel combustion) | Manufacturing industries and construction |
| | Other energy industries |
| Ground Transport | Road transportation |
| | Rail transportation |
| | Inland navigation |
| | Other transportation |
| Residential | Residential and other sectors |

**SI Table S2.** Sectoral changes in the first quarter in 2020 comparing to the same periods in 2019 by countries or regions.

| Emission Decline (MtCO$_2$, 2020F4) | Power | Transport | Industrial (with Process) | Residential | Domestic Aviation | **Sum** | Growth Rates (%) |
|---|---|---|---|---|---|---|---|
| China | -91.1 | -84.4 | -43.8 | -7.5 | -7.8 | **-234.5** | -6.9% |
| India | -39.7 | -21.4 | -22.1 | 7.3 | -0.7 | **-76.6** | -8.5% |
| US | -43.8 | -78.3 | -17.1 | -14.8 | -8.3 | **-162.4** | -9.5% |
| Europe (EU27 & UK) | -82.0 | -26.8 | -15.1 | -13.3 | -1.2 | **-138.3** | -12.0% |
| Russia | -7.0 | -3.6 | 0.2 | -8.1 | -0.3 | **-18.7** | -3.4% |
| Japan | -5.5 | -4.3 | -5.5 | -1.9 | -0.2 | **-17.3** | -4.3% |
| Brazil | 1.1 | -8.3 | -2.2 | 0.0 | -0.5 | **-9.9** | -7.0% |
| Row | -24.6 | -113.1 | -30.7 | -5.1 | -10.3 | **-183.7** | -5.4% |
| **Sum** | **-292.5** | **-340.0** | **-136.2** | **-43.4** | **-29.2** | **-841.4** | **-7.5%** |
| Growth Rates (%) | -6.4% | -15.5% | -4.4% | -2.7% | -23.4% | -7.5% | |
| International aviation | | | | | | -63.6 | -32.4% |
| International shipping | | | | | | -32.6 | 15.0% |
| Global | | | | | | **-937.6** | -7.8% |

**SI Table S3.** Monthly changes in power sector in 2020 comparing to the same periods in 2019 by countries or regions.

| Countries/Regions | Jan | Feb | Mar | Apr | Jan-Apr |
|---|---|---|---|---|---|
| China | -3.6% | -14.4% | -8.0% | 1.1% | -6.0% |
| India | -0.3% | 9.0% | -12.8% | -29.9% | -9.2% |
| US | -10.6% | -2.8% | -8.6% | -8.4% | -7.7% |
| Europe (EU27&UK) | -18.0% | -26.6% | -10.7% | -36.6% | -22.5% |
| Russia | -3.4% | 0.5% | -2.9% | -3.9% | -2.4% |
| Japan | -5.4% | -0.1% | -2.1% | -3.8% | -2.9% |
| Brazil | 98.7% | -10.0% | -18.8% | -24.3% | 6.3% |
| ROW | 0.2% | 0.2% | -1.8% | -7.3% | -2.0% |
| World | -4.3% | -6.4% | -6.5% | -8.9% | -6.4% |

**SI Table S4**. Monthly mobility changes in 2020 comparing to the same periods in 2019 by countries or regions.

| Countries/Regions | Jan | Feb | Mar | Apr | Jan-Apr |
|---|---|---|---|---|---|
| China | -18.5% | -53.4% | -25.9% | -16.3% | -28.1% |
| India | 0.6% | 1.9% | -25.9% | -65.6% | -22.3% |
| US | 11.2% | 9.9% | -22.8% | -50.0% | -13.9% |
| Europe (EU27&UK) | 6.0% | 7.0% | -16.6% | -32.1% | -9.2% |
| Russia | 6.1% | 6.0% | -3.5% | -26.1% | -4.7% |
| Japan | -3.0% | 1.0% | -7.2% | -17.5% | -6.8% |
| Brazil | 3.0% | -0.2% | -15.1% | -37.7% | -12.9% |
| ROW | 3.8% | 0.5% | -22.0% | -42.2% | -15.4% |
| World | 2.5% | -3.3% | -20.9% | -39.0% | -15.5% |

**SI Table S6**. Growth rates of industrial emissions comparing to the same periods of last year in 2020.

| Countries/Regions | Jan | Feb | Mar | Apr | Jan-Apr |
|---|---|---|---|---|---|
| China | -7.5% | -6.2% | -5.0% | 3.3% | -3.5% |
| India | 1.6% | 3.1% | -20.6% | -14.8% | -7.9% |
| US | -0.7% | -0.1% | -5.9% | -18.8% | -6.4% |
| Europe (EU27&UK) | -1.2% | -1.3% | -12.0% | -14.8% | -7.3% |
| Russia | 3.9% | 4.9% | 2.6% | -9.4% | 0.2% |
| Japan | -2.4% | -5.6% | -5.3% | -10.6% | -6.0% |
| Brazil | 1.5% | -0.4% | -4.2% | -16.0% | -5.0% |
| ROW | -2.0% | -2.0% | -3.1% | -6.2% | -3.4% |
| World | -3.4% | -2.8% | -6.1% | -4.8% | -4.4% |

**SI Table S9.** The observation of air quality and dry column $CO_2$ ($XCO_2$) (Full Data file attached separately)

|  |  | China | US | EU4 | India |
|---|---|---|---|---|---|
| **OMI NO2** | Jan | -32.26% ± 12.03% | 22.98% ± 16.02% | 15.78% ± 15.24% | -8.96% ± 13.63% |
|  | Feb | -34.22% ± 11.87% | -23.08% ± 12.63% | -25.12% ± 12.40% | -13.79% ± 13.36% |
|  | Mar | -4.53% ± 13.77% | -14.32% ± 13.24% | -15.56% ± 13.22% | -13.37% ± 13.29% |
|  |  | -3.59% ± 13.97% | -2.16% ± 14.07% | -23.40% ± 12.48% | -11.10% ± 13.50% |
|  | Jan-Mar | -21.55% ± 12.87% | -4.22% ± 13.75% | -12.73% ± 13.26% | -11.78% ± 13.41% |
| **MODIS AOD** | Jan | 10.17% ± 49.22% | 10.64% ± 95.67% | 19.81% ± 77.76% | -5.46% ± 36.27% |
|  | Feb | -7.88% ± 41.08% | -3.98% ± 82.64% | -1.95% ± 70.78% | 7.29% ± 39.89% |
|  | Mar | 3.55% ± 41.88% | -15.65% ± 67.37% | 29.42% ± 62.28% | 1.84% ± 41.12% |
|  | April | 17.35% ± 44.95% | 20.16% ± 85.91% | -23.83% ± 46.15% | -7.65% ± 39.34% |
|  | January-April | 5.34% ± 43.66% | 0.98% ± 81.48% | 2.04% ± 60.55% | -1.06% ± 39.09% |
| **GOSAT XCO2** | Jan | 0.53% ± 0.52% | 0.60% ± 0.52% | 0.42% ± 0.52% | 0.65% ± 0.52% |
|  | Feb | 0.45% ± 0.51% | 0.53% ± 0.52% | 0.65% ± 0.52% | 0.44% ± 0.51% |
|  | Mar | 0.67% ± 0.52% | 0.37% ± 0.52% | 0.51% ± 0.52% | 0.66% ± 0.52% |
|  | April | 0.48% ± 0.51% | 0.46% ± 0.52% | 0.39% ± 0.52% | 0.51% ± 0.52% |
|  | Jan-Mar | 0.53% ± 0.52% | 0.49% ± 0.52% | 0.49% ± 0.52% | 0.56% ± 0.52% |
| **TROPOMI CO** | Jan | 2.67%±5.20% | 4.94%±3.02% | 2.37%±1.89% | -0.41%±3.37% |
|  | Feb | 0.47%±7.11% | 2.97%±3.91% | 1.08%±2.33% | 3.23%±5.20% |
|  | Mar | 3.98%±6.77% | -1.84%±3.39% | -1.14%±3.10% | 2.12%±3.45% |
|  | Jan-Mar | 2.38%±4.84% | 1.85%±1.94% | 0.72%±1.40% | 1.66%±2.38% |
| **Site NO2** | Jan | -18.05%±23.90% | -3.80%±12.80% |  |  |
|  | Feb | -30.33%±21.78% | 14.98%±68.82% |  |  |
|  | Mar | -23.03%±17.29% | -8.98%±44.17% |  |  |
|  | Jan-Mar | -23.00%±14.97% | 0.34%±79.05% |  |  |
| **Site PM2.5** | Jan | -2.67%±41.56% | -8.77%±49.43% |  |  |
|  | Feb | -26.71%±26.94% | -14.78%±59.12% |  |  |
|  | Mar | -21.80%±17.51% | -20.55%±39.30% |  |  |
|  | Jan-Mar | -15.39%±19.06% | -14.68%±40.49% |  |  |
| **Site CO** | Jan | -5.89%±22.22% | -12.35%±23.12% |  |  |
|  | Feb | -19.60%±20.49% | -6.19%±45.39% |  |  |
|  | Mar | -14.24%±19.77% | 4.94%±74.21% |  |  |
|  | Jan-Mar | -12.51%±15.41% | -5.11%±26.53% |  |  |
| **Inventory NO2** | Jan |  | -0.99% (-1.36~-0.86%) |  |  |
|  | Feb |  | 2.43% (2.11~3.33%) |  |  |
|  | Mar | -15.49% (-21.20~-13.46%) | -7.72% (-10.58~-6.72%) |  |  |
|  | Jan-Mar | -17.47% (-23.94~-15.20%) | -2.57% (-3.52~-2.24%) |  |  |